\documentclass{aa}
\usepackage{graphicx}
\usepackage{natbib}
\bibpunct{(}{)}{;}{a}{}{,} %

\usepackage{amsmath}
\usepackage{amssymb} 
\usepackage{wasysym}

\title{Looking for the rainbow on exoplanets covered by liquid and icy water clouds}

\author{T. Karalidi\inst{1,3} \and D. M. Stam\inst{1}
  \and J. W. Hovenier\inst{2}}
        
\institute{
  SRON - Netherlands Institute for Space Research,
  Sorbonnelaan 2, 3584 CA, Utrecht, the Netherlands 
  \and 
  Astronomical
  Institute ``Anton Pannekoek'', University of Amsterdam,
  Science Park 904, 1098 XH Amsterdam, the Netherlands
  \and
  Leiden Observatory, Leiden University, Postbus 9513, 2300 RA, 
  Leiden, the Netherlands}

\date{Received 17-08-2012 / Accepted 23-10-2012}

\begin{document}

\abstract{}{Looking for the primary rainbow in starlight that is
  reflected by exoplanets appears to be a promising method to search
  for liquid water clouds in exoplanetary atmospheres.  Ice water
  clouds, that consist of water crystals instead of water droplets,
  could potentially mask the rainbow feature in the planetary signal
  by covering liquid water clouds.  Here, we investigate the strength
  of the rainbow feature for exoplanets that have liquid and icy water
  clouds in their atmosphere, and calculate the rainbow feature for a
  realistic cloud coverage of Earth.}  {We calculate flux and
  polarization signals of starlight that is reflected by horizontally
  and vertically inhomogeneous Earth--like exoplanets, covered by
  patchy clouds consisting of liquid water droplets or water ice
  crystals. The planetary surfaces are black.}  {On a planet with a
  significant coverage of liquid water clouds only, the total flux
  signal shows a weak rainbow feature. Any coverage of the liquid
  water clouds by ice clouds, however, dampens the rainbow feature in
  the total flux, and thus the discovery of liquid water in the
  atmosphere.  On the other hand, detecting the primary rainbow in the
  polarization signal of exoplanets appears to be a powerful tool for
  detecting liquid water in exoplanetary atmospheres, even when these
  clouds are partially covered by ice clouds.  In particular, liquid
  water clouds covering as little as 10\%--20\% of the planetary
  surface, with more than half of these covered by ice clouds, still
  create a polarized rainbow feature in the planetary signal. Indeed,
  calculations of flux and polarization signals of an exoplanet with a
  realistic Earth--like cloud coverage, show a strong polarized
  rainbow feature.}{}


\titlerunning{Looking for the rainbow of water clouds}
\authorrunning{T. Karalidi et al.}

\maketitle

\section{Introduction}
\label{sect:intro}

The discovery of the first exoplanet orbiting a main sequence star
almost two decades ago \citep{mayorqueloz95} inaugurated a new era in
astronomy. As of today, more than 700 exoplanets have been detected
(source: The extrasolar planets encyclopeadia). Telescope instruments
and satellite missions, like for example COROT (COnvection, ROtation
\& planetary Transits) \citep{baglin06}, NASA's Kepler mission
\citep{koch98}, HARPS (High Accuracy Radial Velocity Planet Searcher)
\citep[e.g.][]{pepe04}, Super--WASP \citep[][]{deming12}, and, in the
near future, GPI (Gemini Planet Imager) \citep[][]{macintosh08} on the
Gemini observatory (first on the telescope on the southern hemisphere)
and SPHERE (Spectro--Polarimetric High--Contrast Exoplanet Research)
\citep[e.g.][]{dohlen08, roelfsema11} on ESO's Very Large Telescope
(VLT), to name a few, will rapidly increase the number of detected
exoplanets.

The detection methods used and the accuracy of our instruments result
in most of the exoplanets detected up to today being giants, even
though in recent years the lower mass limit of our detections has been
pushed down allowing for the detection of more than 30 super--Earth
planets. With an increasing possibility for the detection of the first
Earth--like planets in the next decade, an important factor to
consider is how ready our models will be to interpret the
observations.

An important factor that needs to be taken into account for future
efforts to detect signatures of life on other planets is the possible
inhomogeneity of the planetary surface and atmosphere \citep[][and
  references therein]{tinetti06b}. The existence of continents, oceans
and variable atmospheric patterns (cloud patches etc), as well as
their distributions across the planetary surface can have a large
impact on the observed signal. For this reason, the models that we use
to interpret the observations should be able to handle inhomogeneous
planets.

There exist a number of models that deal with the brightness of
inhomogeneous exoplanets \citep[][to name a few]{ford01, tinetti06a,
  montanesrodrigues06, palle08}. All of the models show the importance
of planetary inhomogeneity and temporal variability on the modelled
planetary signal. The pioneering work of \citet{ford01} as well as
later studies \citep[e.g.][]{oakley09}, show a clear diurnal
variability of the modelled Earth--as--an--exoplanet signal due to
areas of different albedo passing in and out of the observational
field of view.

Among the factors that influence the planetary signal, clouds have a
prominent role. Observations of earthshine for example, have shown
that clouds can induce a considerable daily variation in the planetary
signal. The amount of variability observed differs slightly among the
observations ($\sim10$\% for \citet{palle04}, $\sim5$\% for
\citet{goode01} and a few percent for \citet{cowan09}). Cloud coverage
and variability can also influence to a large degree the
interpretation of the observations. \citet{oakley09} for example find
that mapping the planetary surface is only possible for cloud
coverages smaller than the mean Earth one. Even in the case of giant
planets or dwarf stars, clouds play a crucial role in defining the
atmospheric thermal profile and eventually spectra
\citep[][]{marley10}.

At present, the characterization of exoplanets is mainly done using
planetary transits with instruments on e.g. the Hubble and Spitzer
Space Telescopes \citep[see e.g.][]{ehrenreich07,deming11,tinetti10},
and for some planets, even with ground--based instrumentation
\citep[see e.g.][]{snellen10, brogi12, demooij12}.  With the transit
method though, during the primary transit, the observed starlight has
only penetrated the upper layers of the planetary atmosphere.  Cloud
layers at lower altitudes could for example block out the signal from
lower atmospheric layers or a possible planetary surface. Even with
the help of secondary transits, the characterisation of Earth--like
exoplanets in the habitable zone of a solar--type star would not be
possible, since these planets would yield too weak a signal
\citep[][]{kaltenegger09}.  Direct observations of reflected starlight
from the planet could solve this problem, since then information from
the lower atmospheric layers and surface could survive in the observed
planetary signal. The combination in these cases of flux and
polarization observations could provide us with a crucial tool to
break any possible retrieval degeneracies (for example, such as
between optical thicknesses and single scattering albedo's or cloud
particle sizes) that flux only measurements may present. A first
detection of an exoplanet using polarimetry was claimed by
\citet[][]{berdyugina08}. Subsequent polarization observations by
\citet[][]{wiktorowicz09} could, however, not confirm this
detection. In 2011 \citet[][]{berdyugina11} presented another
detection of the same planet at shorter wavelengths, which still
awaits confirmation by follow--up observations. Telescope instruments
like GPI \citep[][]{macintosh08} and SPHERE \citep[e.g.][]{dohlen08,
  roelfsema11}, which have polarimetric arms that have been optimized
for exoplanet detection, are expected to detect and characterize
exoplanets with polarimetry in the near future.

The power of polarization in studying planetary atmospheres and
surfaces has been shown multiple times in the past through
observations of Solar System planets (including Earth
itself)\citep[see for example][]{hansenhovenier74, hansentravis74,
  mishchenko90, tomasko09}, as well as by modeling of solar system
planets or giant and Earth--like exoplanets \citep[e.~g.~][]{stam03,
  stamhovenier04, saar03, seager00, stam08, karalidi11}. Polarization
provides us with a unique tool for the detection of liquid water on a
planetary atmosphere and surface. \citet[][]{williams08} e.g.~use
polarization to detect the glint of starlight reflected on liquid
surfaces (oceans) of exoplanets and \citet[][]{zugger10, zugger11}
conclude that the existence of an atmosphere as thick as Earth's would
hide any polarization signature of the underlying oceans. Probably the
most interesting feature to look for in the polarization signal of
exoplanets is the rainbow.

The rainbow is a direct indication of the presence of liquid water
droplets in a planetary atmosphere \citep[see e.g.][]{bailey07,
  karalidi11}. Its angular position depends strongly on the refractive
index of the scattering particles and slightly on their effective
radius \citep[see][and references therein]{karalidi11}. Its existence,
most pronounced in polarization observations, can be masked by the
existence of ice clouds \citep[][]{goloub00}. The latter are often
located above thick liquid water clouds and can dominate the
appearance of the reflected polarization signal for ice cloud optical
thicknesses larger than 2 \citep[][]{goloub00}.

In \citet{karalidi11}, we calculated flux and polarization signals for
horizontally homogeneous model planets that were covered by liquid
water clouds. The polarization signals of these planets clearly
contained the signature of the rainbow. The polarization signals of
the quasi horizontally inhomogeneous planets (where weighted sums of
horizontally homogeneous planets are used to approximate the signal of
a horizontally inhomogeneous planet) as presented by \citet{stam08}
also show the signature of the rainbow. However, to confirm that
realistically horizontally inhomogeneous planets, with patchy liquid
water clouds, and with patchy liquid water clouds covered by patchy
ice clouds, also show the signature of the rainbow, the algorithm for
horizontally inhomogeneous planets as described in
\citet[][]{karalidi11b} is required.  In this paper we will do exactly
that: using the algorithm of \citet[][]{karalidi11b} to look for the
rainbow on horizontally inhomogeneous planets, that are covered by
different amounts of liquid and icy water clouds.

This paper is organized as follows. In Sect.~\ref{sec:signals}, we
give a short description of polarized light and our radiative transfer
algorithm, and present the model planets and the clouds we use.  In
Sect.~\ref{sect_3}, we investigate the influence of the liquid water
cloud coverage on the strength of the rainbow feature of a planet in
flux and polarization. In Sects.~\ref{sec:mixed_sizes}
and~\ref{sec:h2oice} we investigate the influence of different cloud
layers, respectively with different droplet sizes and with different
thermodynamic phases (liquid or ice), on the strength of the rainbow
feature. An interesting test case for the detection of a rainbow
feature is of course the Earth itself.  In
Sect.~\ref{sect_signs_earth}, we use realistic cloud coverage, optical
thickness and thermodynamic phase data from the MODIS satellite
instrument to investigate whether the rainbow feature of water clouds
would appear in the disk integrated sunlight that is reflected by the
Earth.  Finally, in Sect.~\ref{sect_summ}, we present a summary of our
results and our conclusions.

\section{Calculating flux and polarization signals}\label{sec:signals}

\subsection{Defining flux and polarization}\label{sec:luxpol_def}

We describe starlight that is reflected by a planet by a 
flux vector $\pi\vec{F}$, as follows
\begin{equation}\label{eq:first}
   \pi\vec{F}= \pi\left[\begin{array}{c} 
               F \\ Q \\ U \\ V \end{array}\right],
\end{equation}
where parameter $\pi F$ is the total flux, parameters $\pi Q$ and $\pi
U$ describe the linearly polarized flux and parameter $\pi V$ the
circularly polarized flux \citep[see e.g.][]{hansentravis74,
  hovenier04, stam08}. All four parameters depend on the wavelength
$\lambda$, and their dimensions are W~m$^{-2}$m$^{-1}$. Parameters
$\pi Q$ and $\pi U$ are defined with respect to a reference plane, and
as such we chose here the planetary scattering plane, i.e. the plane
through the center of the star, the planet and the observer.
Parameter $\pi V$ is usually small \citep[][]{hansentravis74}, and we
will ignore it in our numerical simulations. Ignoring $\pi V$ will not
introduce significant errors in our calculated total and polarized
fluxes \citep{stam05}.

The linearly polarized flux, $\pi F_\mathrm{P}$ of flux
vector $\pi \vec{F}$ is independent of the choice of the reference
plane and is given by
\begin{equation}\label{eq:polflux}
   \pi F_{\rm P} = \pi \sqrt{Q^2 + U^2},
\end{equation}
while the degree of (linear) polarization $P$ is defined as the ratio
of the linearly polarized flux to the total flux, as follows
\begin{equation}\label{eq:poldef}
   P= \frac{F_{\rm P}}{F} = \frac{\sqrt{Q^{2}+U^{2}}}{F}.
\end{equation}
For a planet that is mirror--symmetric with respect to the planetary
scattering plane, parameter $U$ will be zero.  In this case we use the
\textit{signed} degree of linear polarization $P_\mathrm{s}= -Q/F$,
which includes the direction of polarization: if $P_\mathrm{s} > 0$
($P_\mathrm{s} < 0$), the light is polarized perpendicular (parallel)
to the plane containing the incident and scattered beams of light.  A
planet with patchy clouds will usually not be mirror--symmetric with
respect to the planetary scattering plane, and hence $U$ will usually
not equal zero.

The flux vector $\pi \vec{F}$ of stellar light that has been reflected
by a spherical planet with radius $r$ at a distance $d$ from the
observer ($d \gg r$) is given by \citep{stam06}
\begin{equation} \label{eq:fluxSF}
   \pi \vec{F}(\lambda,\alpha) =
   \frac{1}{4}\frac{r^2}{d^2}\vec{S}(\lambda,\alpha) \pi \vec{F}_0(\lambda).
\end{equation}
Here, $\lambda$ is the wavelength of the light and $\alpha$ the
planetary phase angle, i.~e.\ the angle between the star and the
observer as seen from the center of the planet. Furthermore, $\vec{S}$
is the 4$\times$4 planetary scattering matrix \citep{stam06} with
elements $a_{ij}$ and $\pi \vec{F}_0$ is the flux vector of the
incident stellar light.  For a solar type star, the stellar flux can
be considered to be unpolarized when integrated over the stellar disk
\citep{kemp87}.

We assume that the ratio of the planetary radius $r$ and the distance
to the observer $d$ is equal to one, and that the incident stellar
flux $\pi F_0$ is equal to $1$~W~m$^{-2}$~m$^{-1}$.  The hence
normalised flux $\pi F_{\mathrm{n}}$ that is reflected by a planet is
thus given by
\begin{equation}
   \pi F_{\mathrm{n}}(\lambda,\alpha) = \frac{1}{4} a_{11}(\lambda,\alpha)
\end{equation}
\citep[see][]{stam08,karalidi11}, and corresponds to the planet's
geometric albedo $A_{\mathrm{G}}$ when $\alpha=0^\circ$. The
corresponding normalized polarized flux is given by 
\begin{equation}
   \pi F_{\mathrm{n,P}}(\lambda,\alpha) = \frac{1}{4}
     \sqrt{a_{21}(\lambda,\alpha)^2 + a_{31}(\lambda,\alpha)^2},
\end{equation}
with $a_{21}$ and $a_{31}$ elements of the planetary scattering
matrix.  Our normalized fluxes $\pi F_{\mathrm{n}}$ can
straightforwardly be scaled for any given planetary system using
Eq.~\ref{eq:fluxSF} and inserting the appropriate values for $r$, $d$
and $\pi F_0$. The degree of polarization $P$ is independent of $r$,
$d$ and $\pi F_0$, and will thus not require any scaling.

\subsection{The radiative transfer calculations}\label{sec:the code}

The code that we use to calculate the total and polarized fluxes of
starlight that is reflected by a model planet fully includes single
and multiple scattering, and polarization. It is based on the same
efficient adding-doubling algorithm \citep[][]{dehaan87} used by
\citet{stam06,stam08}.  To calculate flux and polarization signals of
horizontally inhomogeneous planets, we divide a model planet in pixels
with a size such that we can assume that the surface and atmospheric
layers are locally plane parallel and horizontally homogeneous. We
then use the code for horizontally inhomogeneous planets, as presented
in \citet{karalidi11b}: the contribution of every illuminated pixel
that is visible to the observer to the planet's total and polarized
flux are calculated separately, the polarized fluxes are rotated to
the common reference plane, and then all fluxes are summed up to get
the disk--integrated planetary total and polarized fluxes. From these
fluxes, the disk--integrated degree of linear polarization $P$ is
derived (Eq.~\ref{eq:poldef}).

As before \citep{karalidi11b}, we divide our model planets in
plane--parallel, horizontally homogeneous (but vertically
inhomogeneous) pixels of $2^\circ\times2^\circ$.  Our pixels are large
enough to be able to ignore adjacency effects, i.e. light that is
scattered and/or reflected within more than one pixel (e.g. light that
is reflected by clouds in one pixel towards the surface of another
pixel).  These effects, that make the fluxes emerging from a given
type of pixel dependent on the properties of the surrounding pixels,
show up for higher spatial resolutions, for example, with pixels that
are smaller than about 1$\times$1~km$^2$ \citep{marshak08}.

\subsection{The model planets}
\label{sect_2.3}

\begin{figure}
\centering
\includegraphics[width=85mm]{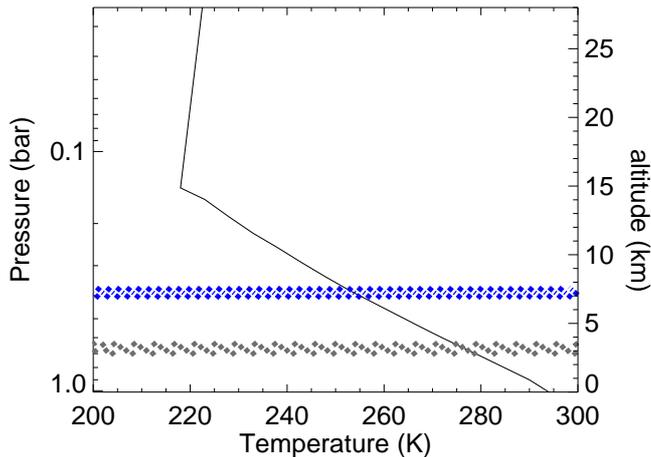}
\caption{The atmospheric pressure and altitude as functions of the
  temperature for the model planets. The liquid water clouds are
  located below 4~km (for example in between 3~km and 4~km as in the
  gray, dotted area), and the ice clouds between 7~km and 8~km (blue,
  dashed--dotted area).}
\label{fig:profiles}
\end{figure}

All model planets have a vertical inhomogeneous atmosphere on top of a
black surface.  The assumption of a black surface is a very good
approximation for an ocean surface (without the glint), which covers
most of the Earth's surface. We'll discuss the effects of brighter
surfaces on our results where necessary.  All model atmospheres have a
pressure and temperature profile representative for a mid-latitude
atmosphere \citep{mcclatchey72} (see Fig.~\ref{fig:profiles}). We
divide each model atmosphere in the same 16 layers. The total gaseous
(Rayleigh) scattering optical thickness of the atmosphere is 0.097 at
$\lambda=0.550~\mu$m and 0.016 at $\lambda=0.865~\mu$m.  We do not
include gaseous absorption, which is a good assumption at the
wavelengths of our interest.

The flux and polarization signals of our model planets are fairly
insensitive to the pressure and temperature profiles, but they are
sensitive to the horizontal and vertical distribution of the clouds,
and to the microphysical properties of the cloud particles.  We will
use two types of clouds: the first type consisting of liquid water
droplets, and the second type consisting of water ice particles. The
details of these particles will be discussed in
Sect.~\ref{sect_particles}.  The liquid water clouds will be located
below an altitude of 4~km (i.e. at pressures lower than 0.628~bars,
which corresponds to temperatures higher than 273~K) and the ice
clouds above that altitude.

\begin{figure}
\centering
\includegraphics[width=85mm]{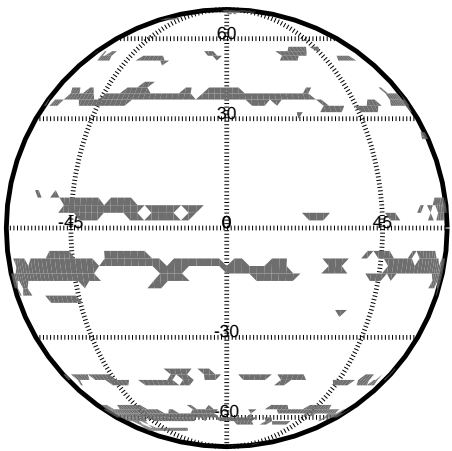}
\caption{A sample map of our model planets. Here, the planetary
  atmosphere contains a single layer with clouds that cover
  $\sim$~25\% of the planet (the dark regions are the clouds).}
\label{fig:cloudmap1}
\end{figure}

We create our cloud maps by using an ISCCP yearly cloud map which we
filter according to the clouds' optical thicknesses.
Figure~\ref{fig:cloudmap1} shows a sample map for a model planet with
a single cloud layer that covers $\sim$25\% of the planet. As shown in
\citet{karalidi11b}, the precise location of clouds can influence in
particular the polarized phase function of a planet (i.e. the degree
of polarization of the reflected starlight as a function of the
planetary phase angle). To avoid changes in $\pi F_\mathrm{n,P}$ and
$P$ due to the location of clouds when the coverage increases, we
increase the cloud coverage of a planet by letting existing clouds
grow in size.  Cloudy regions thus remain cloudy, while surface
patches are getting cloudy.

\subsection{The cloud particles}
\label{sect_particles}

The liquid water cloud particles are spherical, with the standard size
distribution given by \citep[][]{hansentravis74}
\begin{equation}\label{eq:size}
   n(r) = C r^{(1-3 v_{\rm eff})/v_{\rm eff}} 
          e^{-r/ r_{\rm eff} v_{\rm eff}},
\end{equation}
with $n(r) \mathrm{d}r$ the number of particles per unit volume with
radius between $r$ and $r + \mathrm{d}r$, $C$ a constant of
normalization, $r_\mathrm{eff}$ the effective radius and
$v_\mathrm{eff}$ the effective variance of the distribution.
Terrestrial liquid water clouds are in general composed of droplets
with radii ranging from $\sim 5~\mu$m to $\sim 30~\mu$m \citep{han94}.
We use either small droplets (type~A), with $r_\mathrm{eff}=
2.0$~$\mu$m and $v_\mathrm{eff}= 0.1$, or larger droplets (type~B)
with $r_\mathrm{eff}= 6.0$~$\mu$m and $v_\mathrm{eff}= 0.4$ \citep[see
  also][]{karalidi11}. The latter are similar to those used by
\citet{vdiedenhoven} as average values for terrestrial water
clouds. We use a wavelength independent refractive index of
1.335+0.00001$i$ \citep[see][and references therein]{karalidi11}. For
a given wavelength, and values of $r_\mathrm{eff}$ and
$v_\mathrm{eff}$, we calculate the extinction cross--section, single
scattering albedo and the single scattering matrix using Mie--theory
\citep[][]{derooijvdstap87} and normalized according to Eq.~2.5 of
\citet[][]{hansentravis74}.

Ice crystals in nature present a large variety of shapes, depending on
the temperature and humidity conditions during their growth
\citep[][]{hess98}. Even though the variety of shapes is large,
\citet[][]{magono66} showed that only a small number of classes
suffice for the categorization of most natural ice crystals. Until
recently most researchers used perfect hexagonal columns and plates in
order to model the light scattered by natural ice crystals, including
halo phenomena. However, when these particles are oriented, they give
rise to halos, which are rarely observed
\citep[][]{macke96,hess98}. For this reason a number of models were
created to model the signal of ice crystals without halo phenomena
\citep[][]{hess94,macke96,hess98}. In this paper we use an updated
version of the model crystals of \citet[][]{hess98}. Finally, we
should mention that the sizes of most atmospheric ice crystals are
considerably larger than the visible wavelengths we will use in this
paper \citep[][]{macke96}. In particular, the crystals we use in this
paper range in size from $\sim$6 $\mu$m up to $\sim$2 mm, with a size
distribution based on \citep[][]{heymsfield84}.


\begin{figure}[h!]
\centering
\includegraphics[width=85mm]{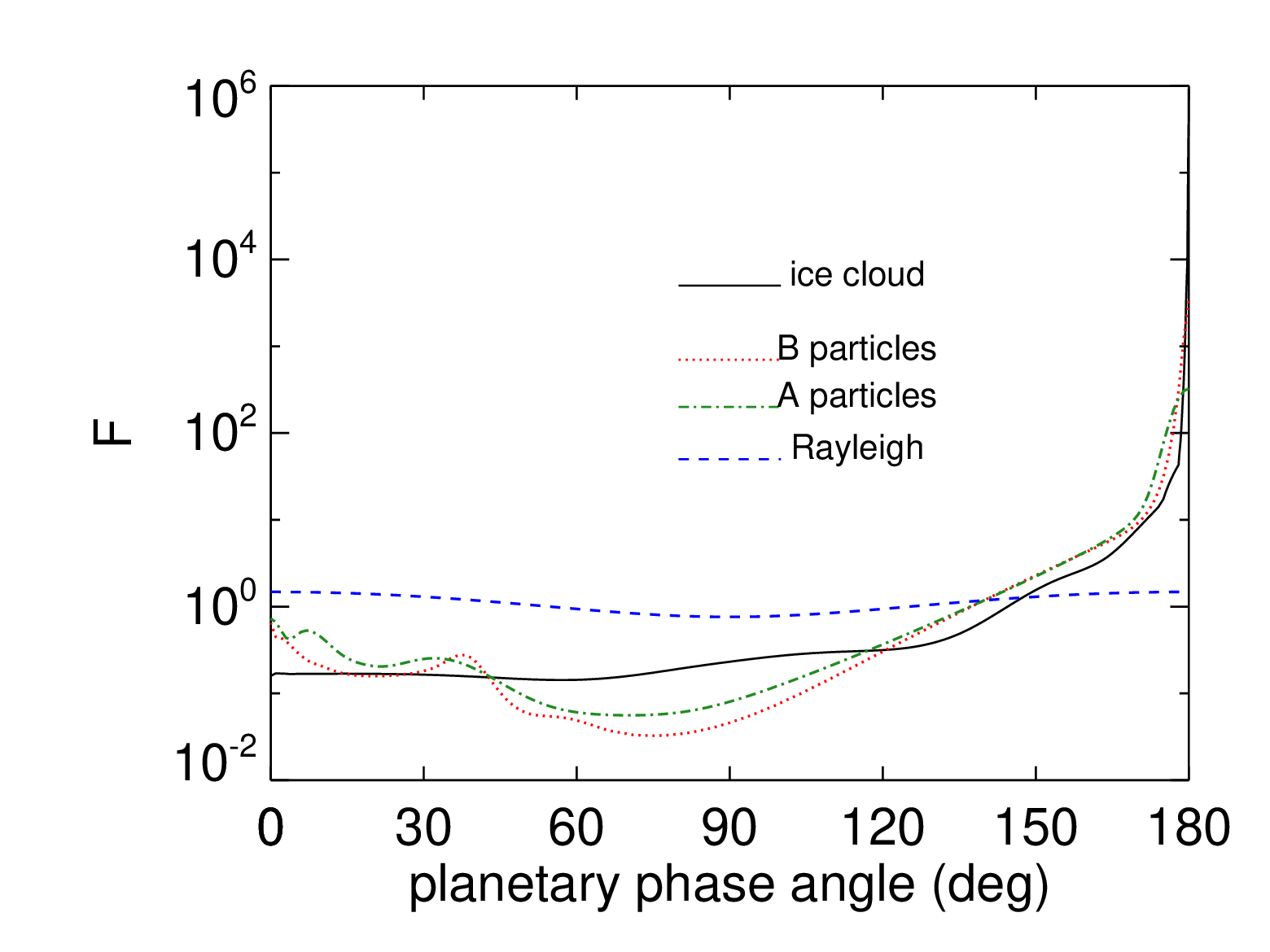}
\hspace{0.8cm}
\centering
\includegraphics[width=85mm]{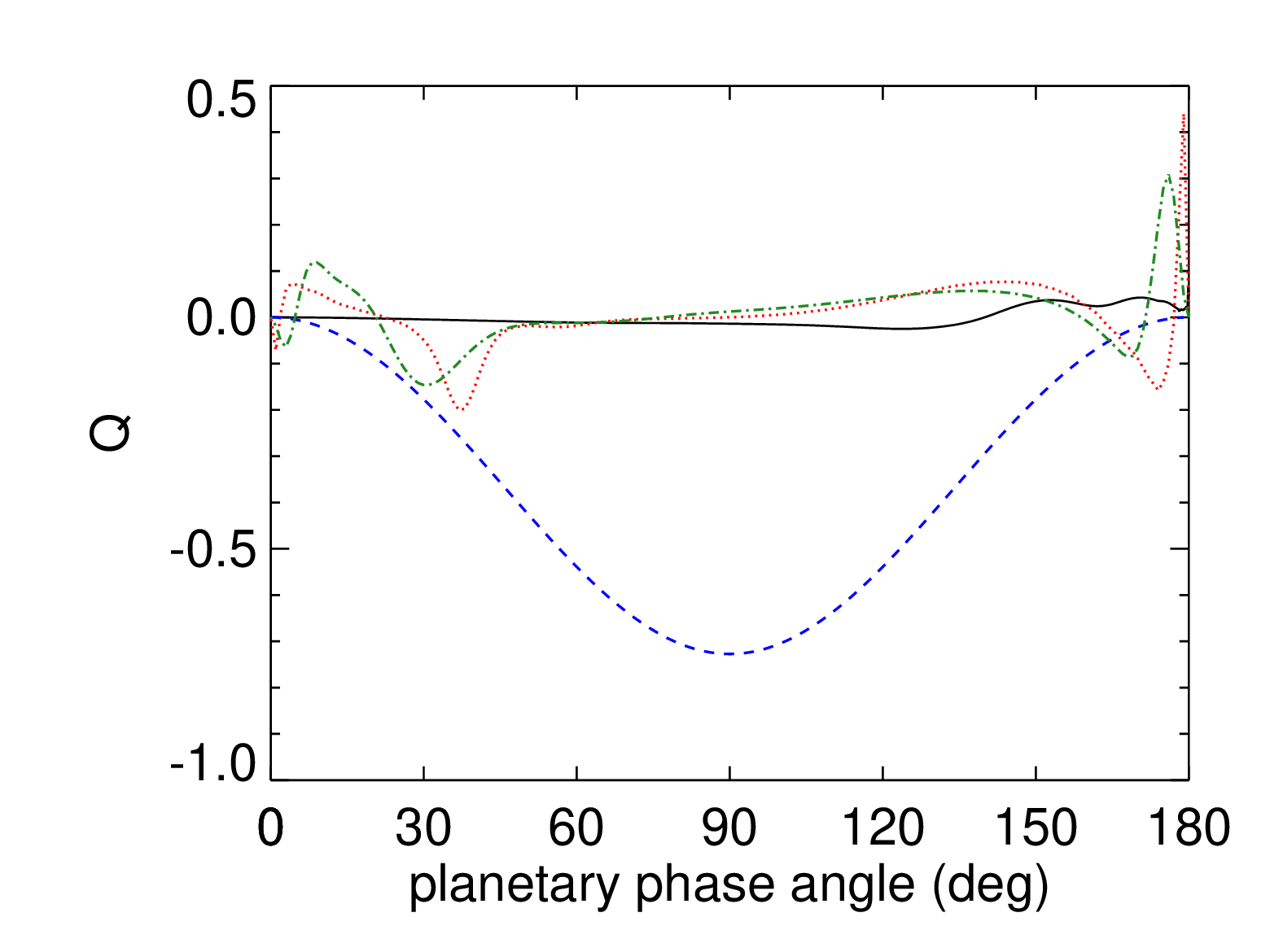}
\hspace{0.8cm}
\centering
\includegraphics[width=85mm]{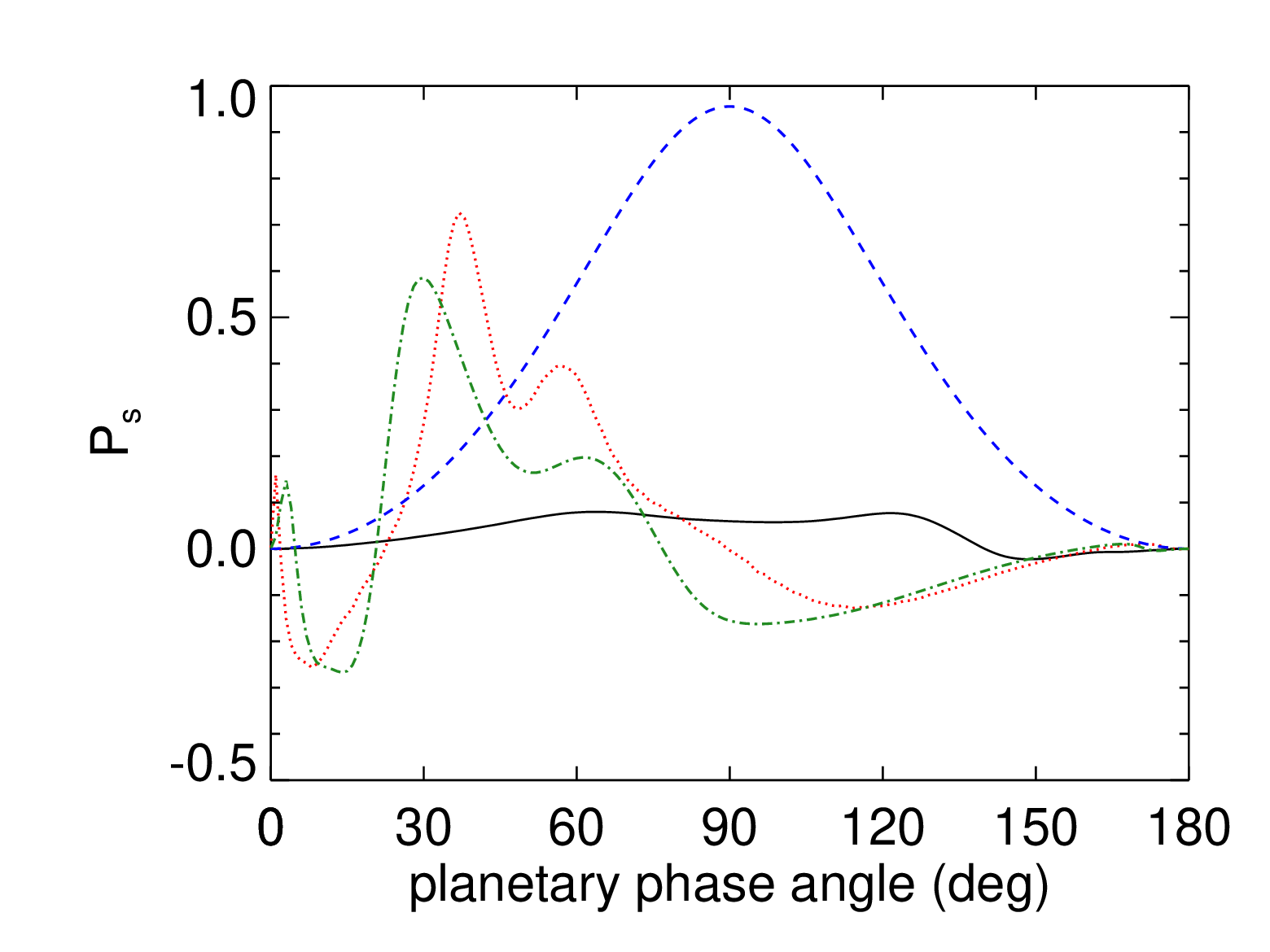}
\caption{Single scattering phase function $F$, linearly polarized flux
  $Q$, and degree of linear polarization $P_\mathrm{s}$ as functions
  of the phase angle $\alpha$ ($\alpha= 180^\circ - \Theta$, with
  $\Theta$ the single scattering angle). Linearly polarized flux $U$
  equals zero, and $P_{\rm s}$ is defined as $-Q/F$.  Curves are shown
  for liquid water droplets type~A ($r_{\rm eff}=0.2~\mu$m, $v_{\rm
    eff}=0.1$), type~B ($r_{\rm eff}=6.0~\mu$m, $v_{\rm eff}=0.4$),
  and water ice particles \citep[][]{hess94,hess98}, all calculated at
  $\lambda=0.550~\mu$m.  For comparison, the curves for Rayleigh
  scattered light are also shown.}
\label{fig:sing_scat_ice}
\end{figure}

Figure~\ref{fig:sing_scat_ice} shows the total flux, the polarized
flux and the degree of linear polarization of light that has been
singly scattered by the cloud droplets and the ice particles at a
wavelength $\lambda$ equal to 0.550~$\mu$m, when the incident light is
unpolarized.  The curves for scattering by gas molecules (Rayleigh
scattering) have also been added.  All curves have been plotted as
functions of the phase angle $\alpha$ instead of of the more common
single scattering angle $\Theta$ to facilitate the comparison with the
signals of the planets. Since for single scattering by the spherical
cloud particles or by the ensemble of randomly oriented ice crystals,
the scattered light is oriented parallel or perpendicular to the
scattering plane (the plane containing the directions of propagation
of the incident and the scattered light), we use the {\em signed}
degree of linear polarization $P_{\rm s}$ in
Fig.~\ref{fig:sing_scat_ice}: when $P_{\rm s}$ is larger (smaller)
than 0, the direction of polarization is perpendicular (parallel) to
the scattering plane. The absolute value of $P_{\rm s}$ equals $P$.

The curves for Rayleigh scattering by gas clearly show the nearly
isotropic scattering of the total flux, the symmetry of the
polarization phase function and the high polarization values around
$\alpha=90^\circ$.  The curves for the total flux scattered by both
types of liquid water droplets have a strong forward scattering peak
at the largest values of $\alpha$ (the smallest single scattering
angles), which is due to refraction and depends mostly on the size of
the scattering particles, and not so much on their shape
\citep[][]{mishchenko10}.  Another characteristic of the flux phase
function of the droplets is the primary rainbow, that is due to light
that has been reflected once inside the particles. For a given
wavelength, the precise location of the rainbow depends on the
particle size: for the small type~A particles, it is found close to
$\alpha=30^\circ$, while for the larger type~B particles, it is close
to $\alpha=40^\circ$.  The primary rainbow also shows up in the
scattered polarized flux and very strongly in the degree of
polarization $P_\mathrm{s}$. The direction of polarization across the
primary rainbow is perpendicular to the planetary scattering plane.
The polarized flux and $P_\mathrm{s}$ go through zero, thus change
direction, a few times between phase angles of 0$^\circ$ and
180$^\circ$. These particular phase angles are usually referred to as
{\em neutral points} and, like the rainbow, they depend on the
particle properties and the wavelength.  At 0.550~$\mu$m, the type~A
particles have neutral points at 5$^\circ$, 20$^\circ$, 76$^\circ$ and
158$^\circ$, and the type~B particles at 2$^\circ$, 22$^\circ$,
94$^\circ$, and 160$^\circ$.

The total flux scattered by the ice particles \citep[][]{hess94,
  hess98} has a smooth appearance without a feature such as the
rainbow but with a strong forward scattering peak at the largest
values of $\alpha$.  The polarized flux and the degree of polarization
of the ice particles are also smooth functions of $\alpha$.  
The neutral point of the ice particles is around $\alpha=140^\circ$.


\begin{figure}
\centering
\includegraphics[width=85mm]{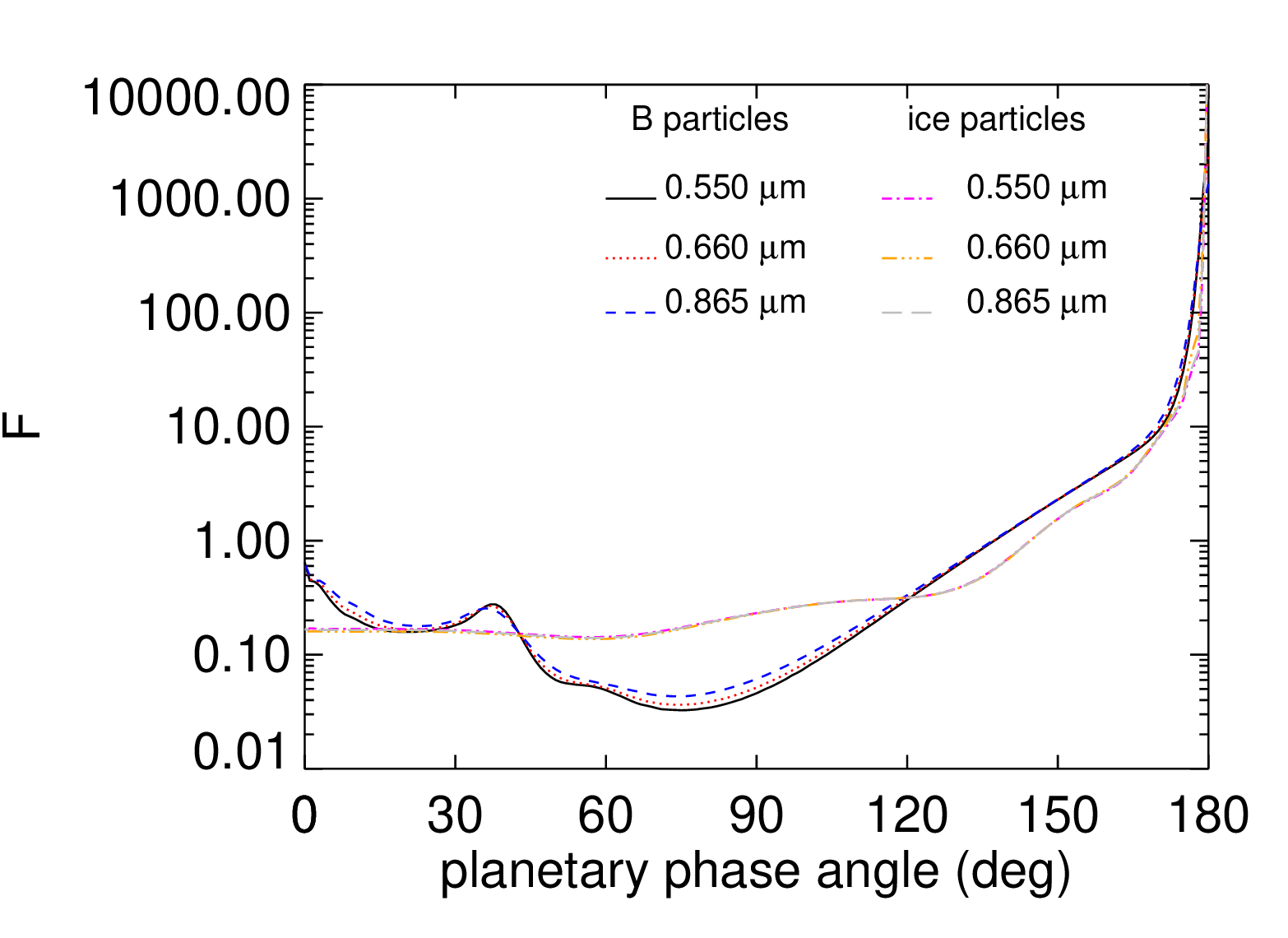}
\hspace{0.8cm}
\centering
\includegraphics[width=85mm]{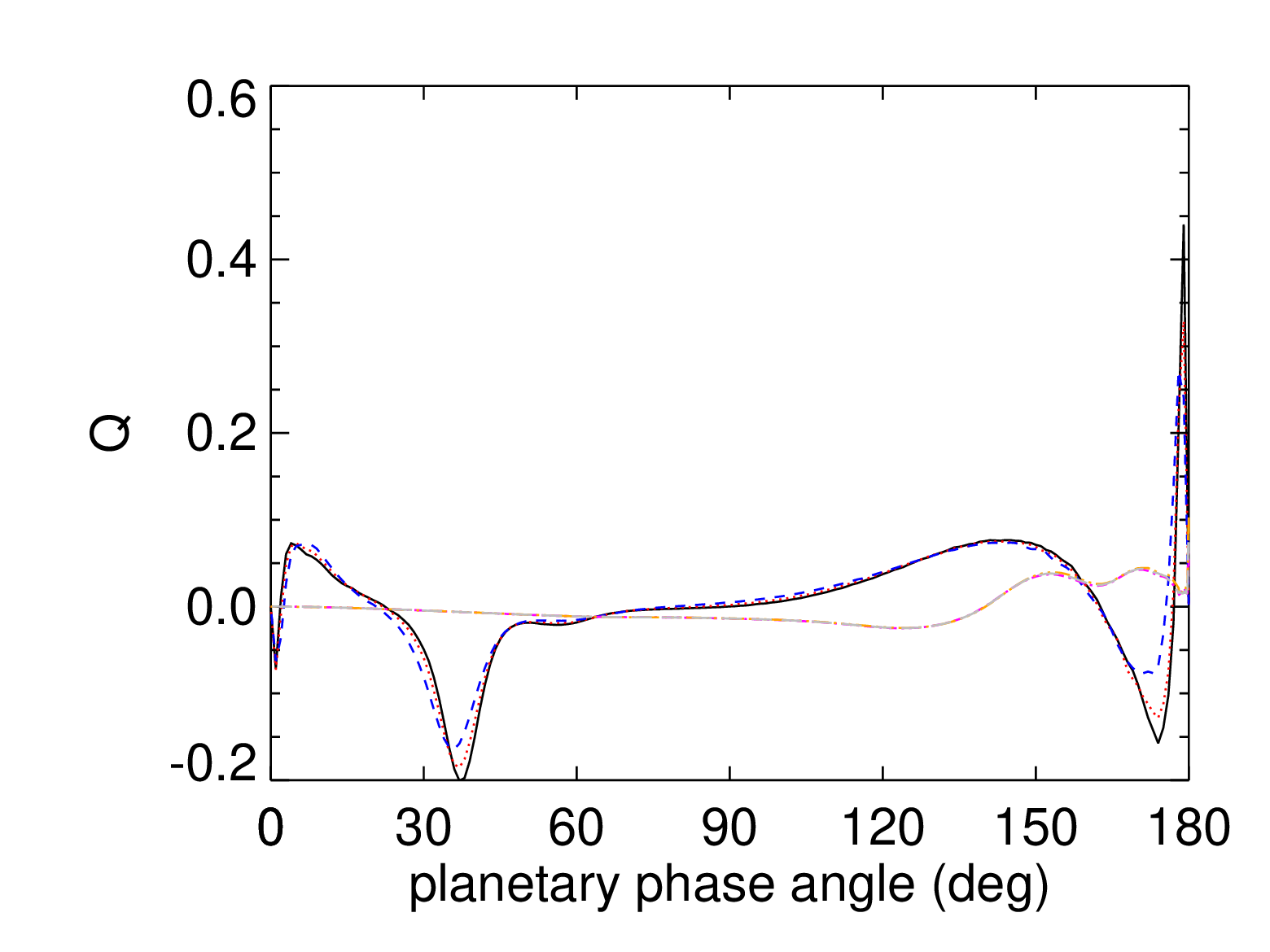}
\hspace{0.8cm}
\centering
\includegraphics[width=85mm]{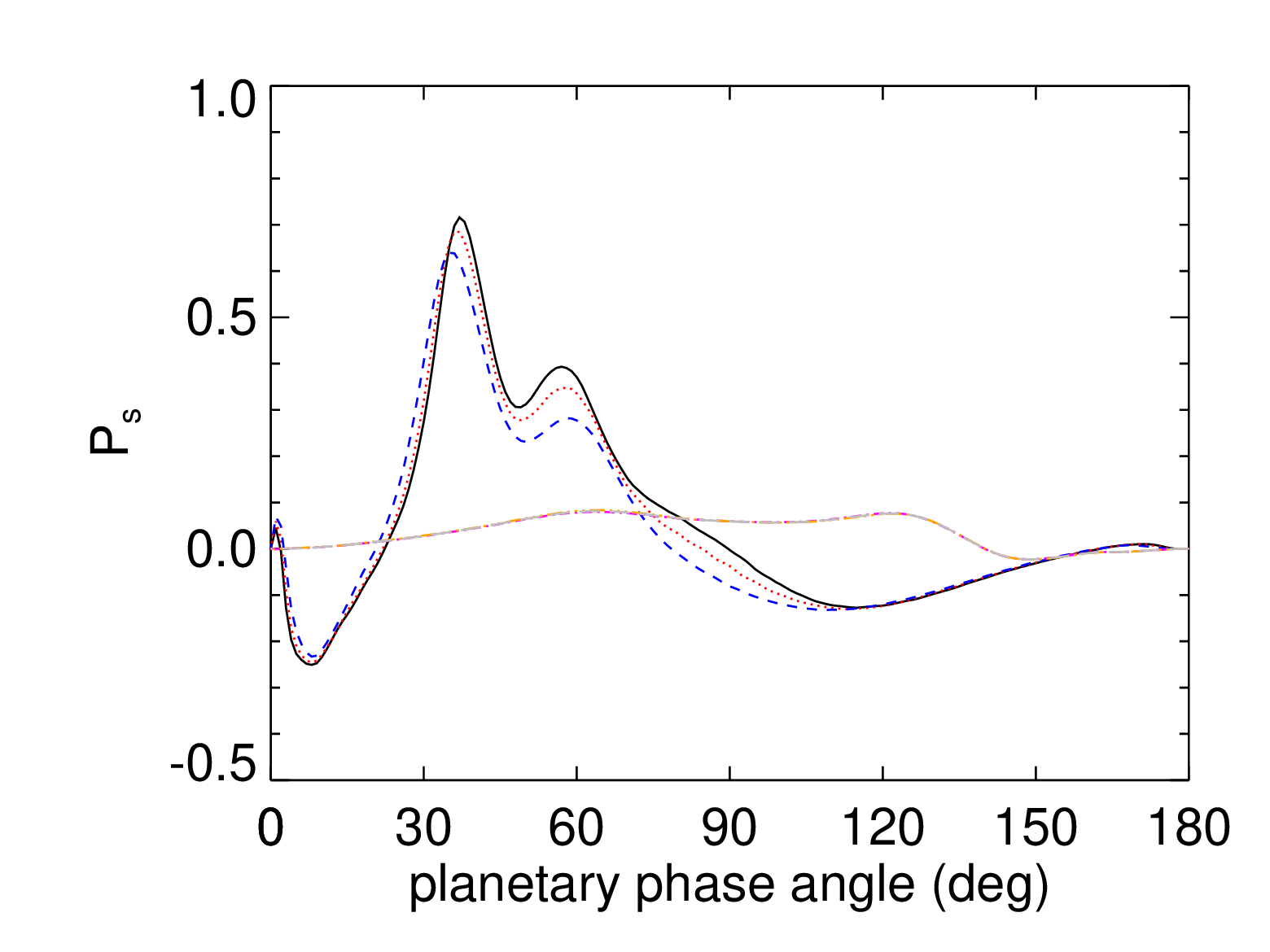}
\caption{Singly scattered $F$, $Q$, and $P_\mathrm{s}$ as functions of
  $\alpha$ for the liquid water droplets type~B and the ice particles
  at $\lambda= 0.550$, $0.660$ and 0.865~$\mu$m.  The curves for the
  ice particles overlap each other.}
\label{fig:sing_scat_ice_wavs}
\end{figure}

Fig.~\ref{fig:sing_scat_ice_wavs} shows the degree of linear
polarization $P_\mathrm{s}$ for the type~B liquid water droplets and
the ice particles at $\lambda=0.550$, $0.660$ and 0.865~$\mu$m.  The
ice particles are relatively large, and therefore their scattering
properties are virtually insensitive to $\lambda$ across the
wavelength region of our interest (the visible).  The scattering
properties of the liquid cloud particles vary slightly with the
wavelength: with increasing $\lambda$, the strengths of the primary
and secondary rainbows decreases slightly, and the neutral point
around intermediate phase angles shifts towards smaller phase angles.
With increasing $\lambda$, the primary rainbow also shifts towards
smaller phase angles (larger single scattering angles). As also
discussed in \citet{karalidi11}, the dispersion of the primary rainbow
depends on the size of the particles that scatter the incident
starlight: in large particles, such as rain droplets, the dispersion
shows the opposite behavior, with the primary rainbow shifting towards
larger phase angles with increasing $\lambda$. This latter shift gives
rise to the well--known primary rainbow seen in the rainy sky with the
red bow on top and the violet bow at the bottom.

\section{The influence of the liquid water cloud coverage}
\label{sect_3}

In this section, we explore the strength of the primary rainbow
feature as a function of the liquid water cloud coverage.  All clouds
have the same optical thickness, i.e. 2.0 (at 0.550~$\mu$m), and the
same altitudes of their bottoms and tops, i.e. 3 and 4~km,
respectively (see Fig.~\ref{fig:profiles}).  The clouds consist of
type~A droplets.

%

\begin{figure}
\centering
\includegraphics[width=85mm]{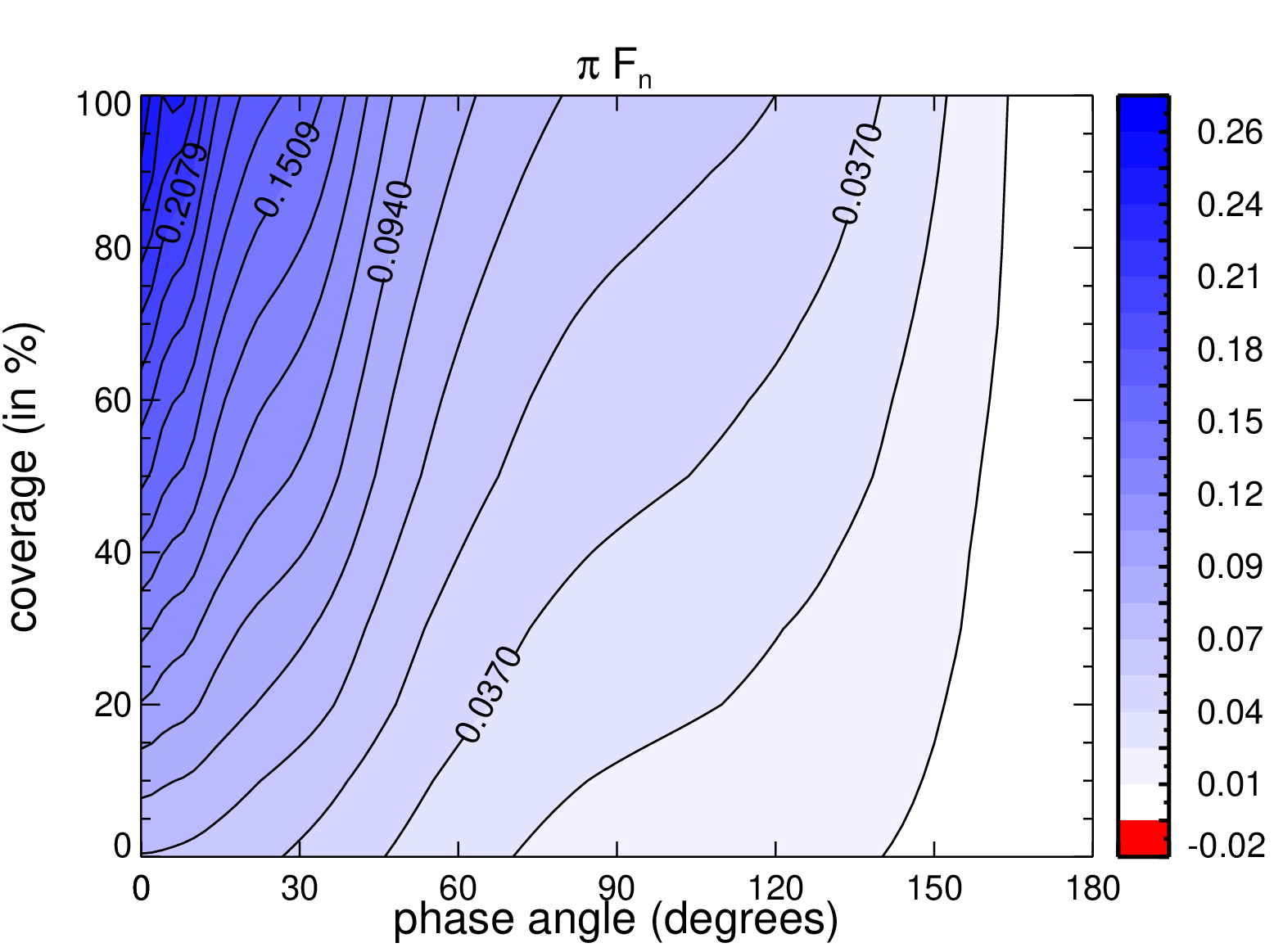}
\hspace{0.8cm}
\centering
\includegraphics[width=85mm]{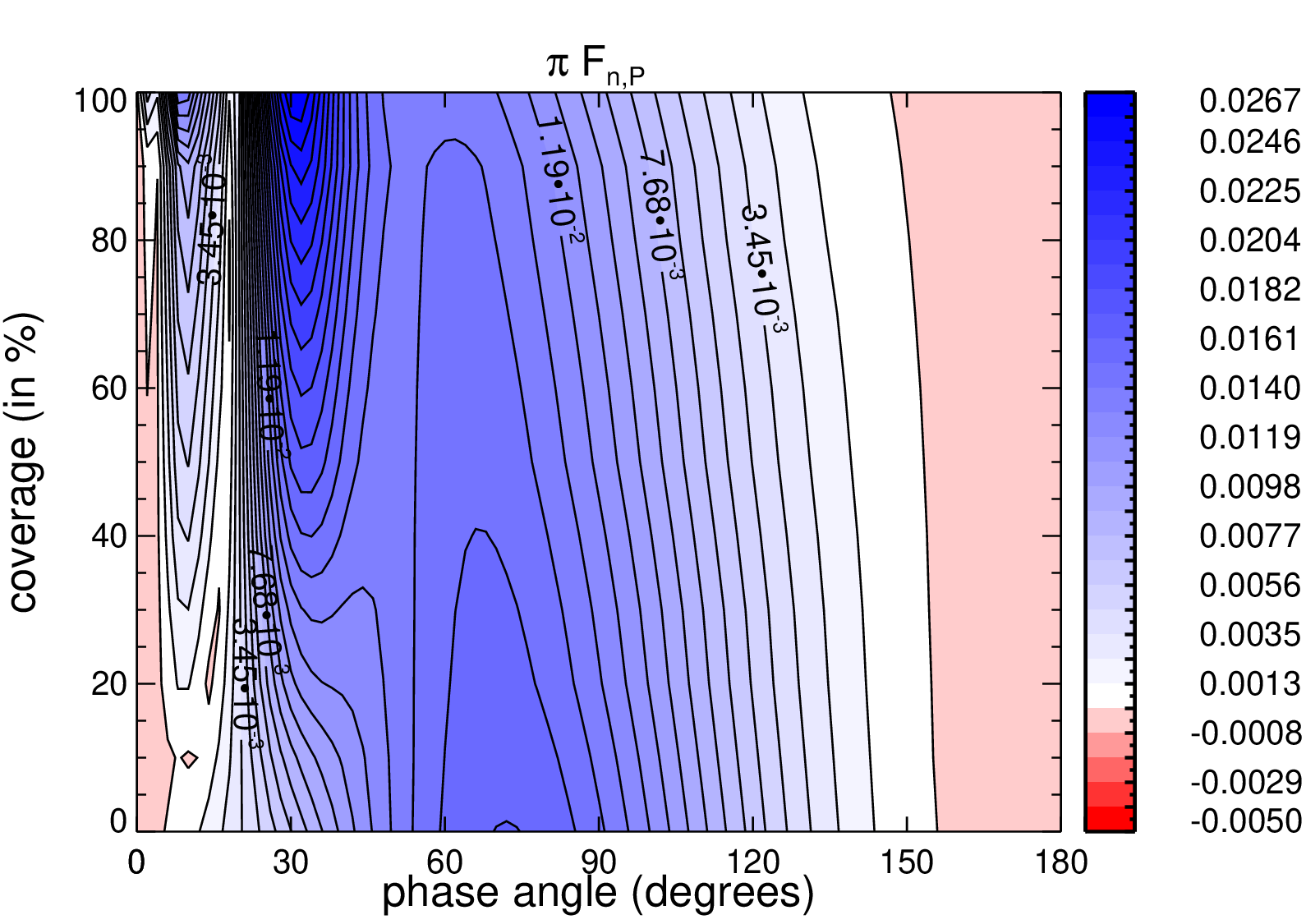}
\hspace{0.8cm}
\centering
\includegraphics[width=85mm]{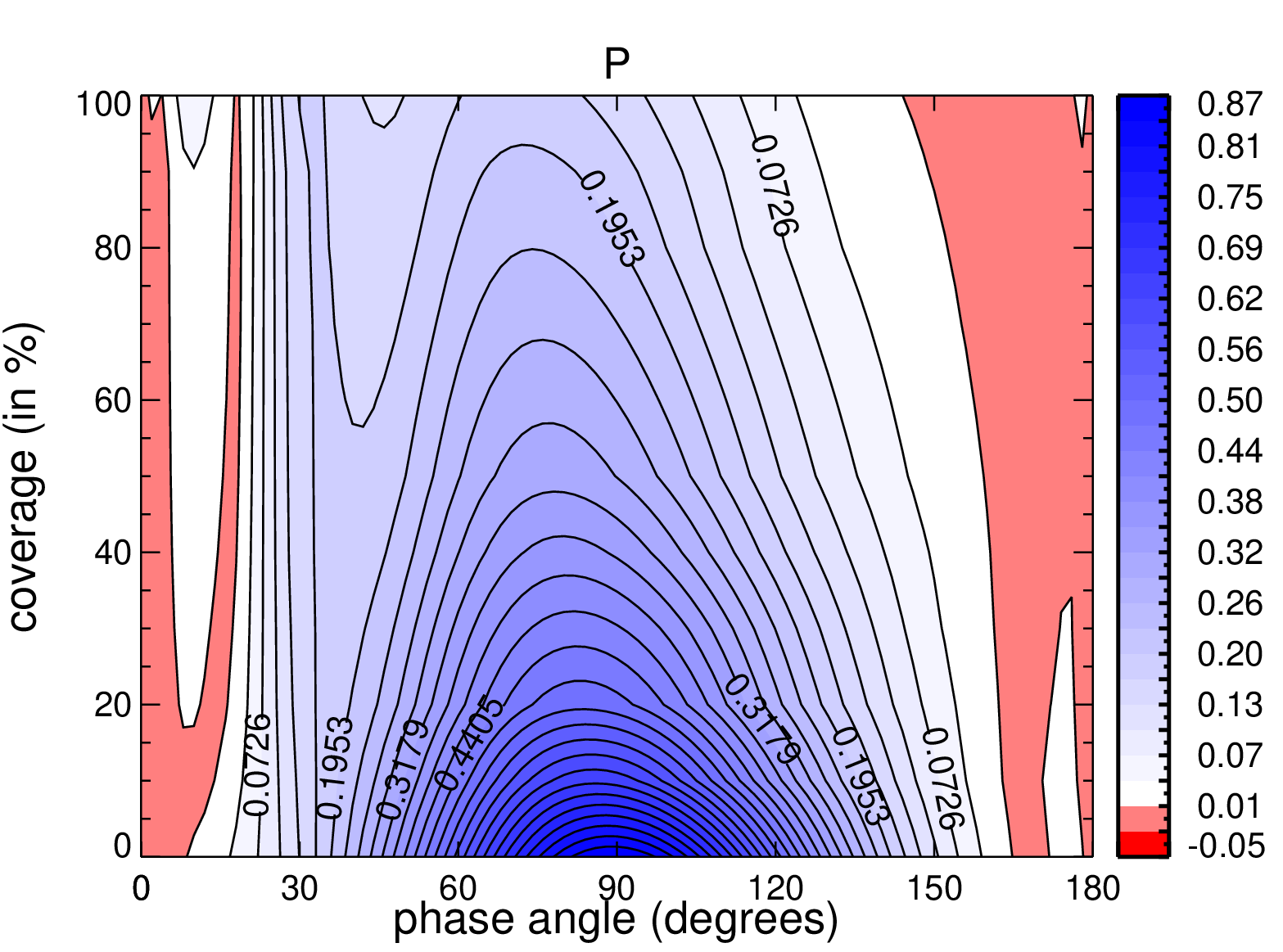}
\caption{The total flux $\pi F_\mathrm{n}$, the polarized flux $\pi
  F_\mathrm{n,P}$, and the degree of polarization $P$ as functions of
  the phase angle $\alpha$ for liquid water cloud coverages from 0\%
  up to 100\%.  Wavelength $\lambda$ is 0.550~$\mu$m, the optical
  thickness of the clouds is 2.0, and their top is at 4~km.  The
  clouds are composed of type~A droplets.}
\label{fig:all_clcv_fp}
\end{figure}

\begin{figure}
\centering
\includegraphics[width=85mm]{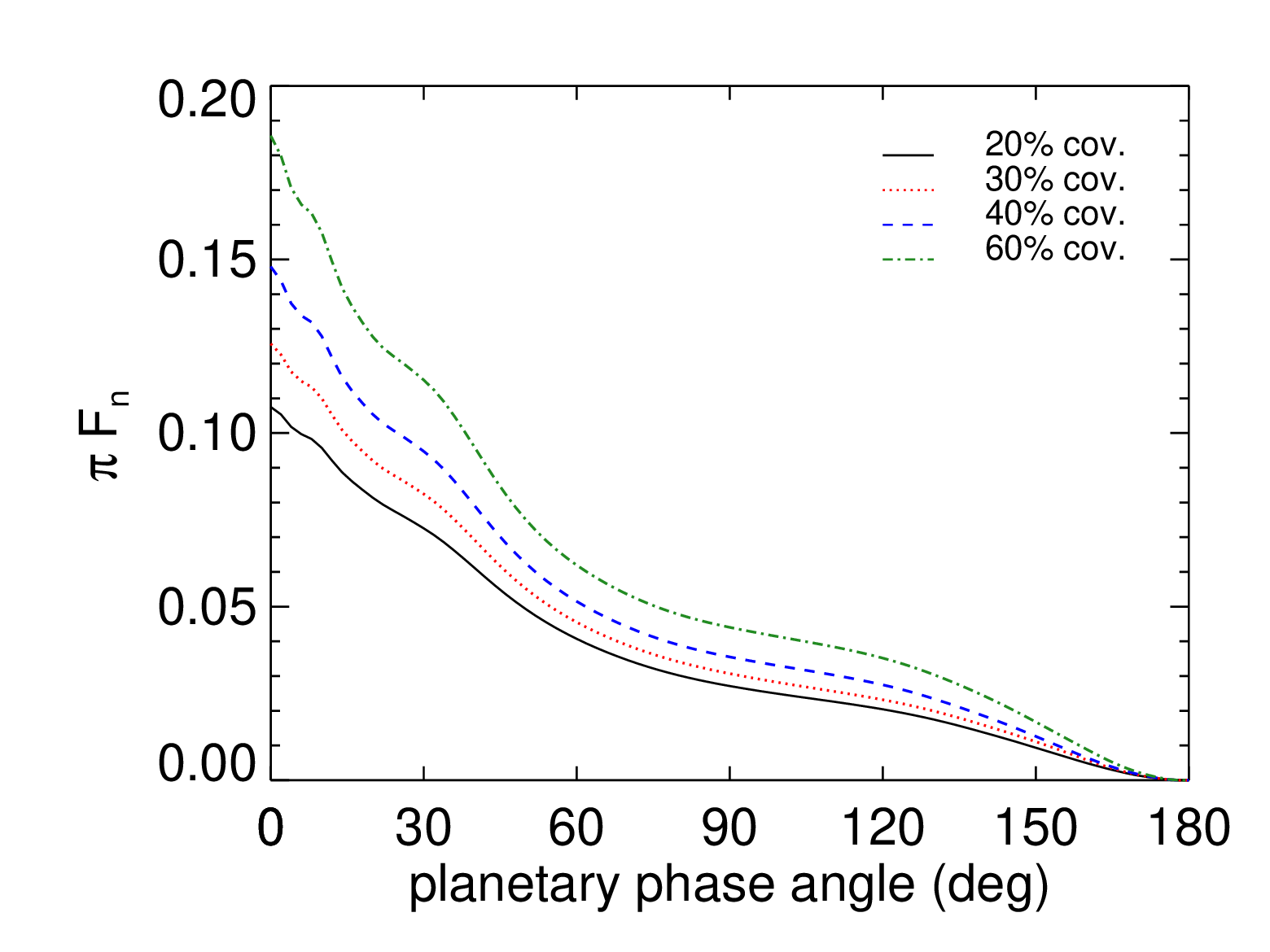}
\hspace{0.8cm}
\centering
\includegraphics[width=85mm]{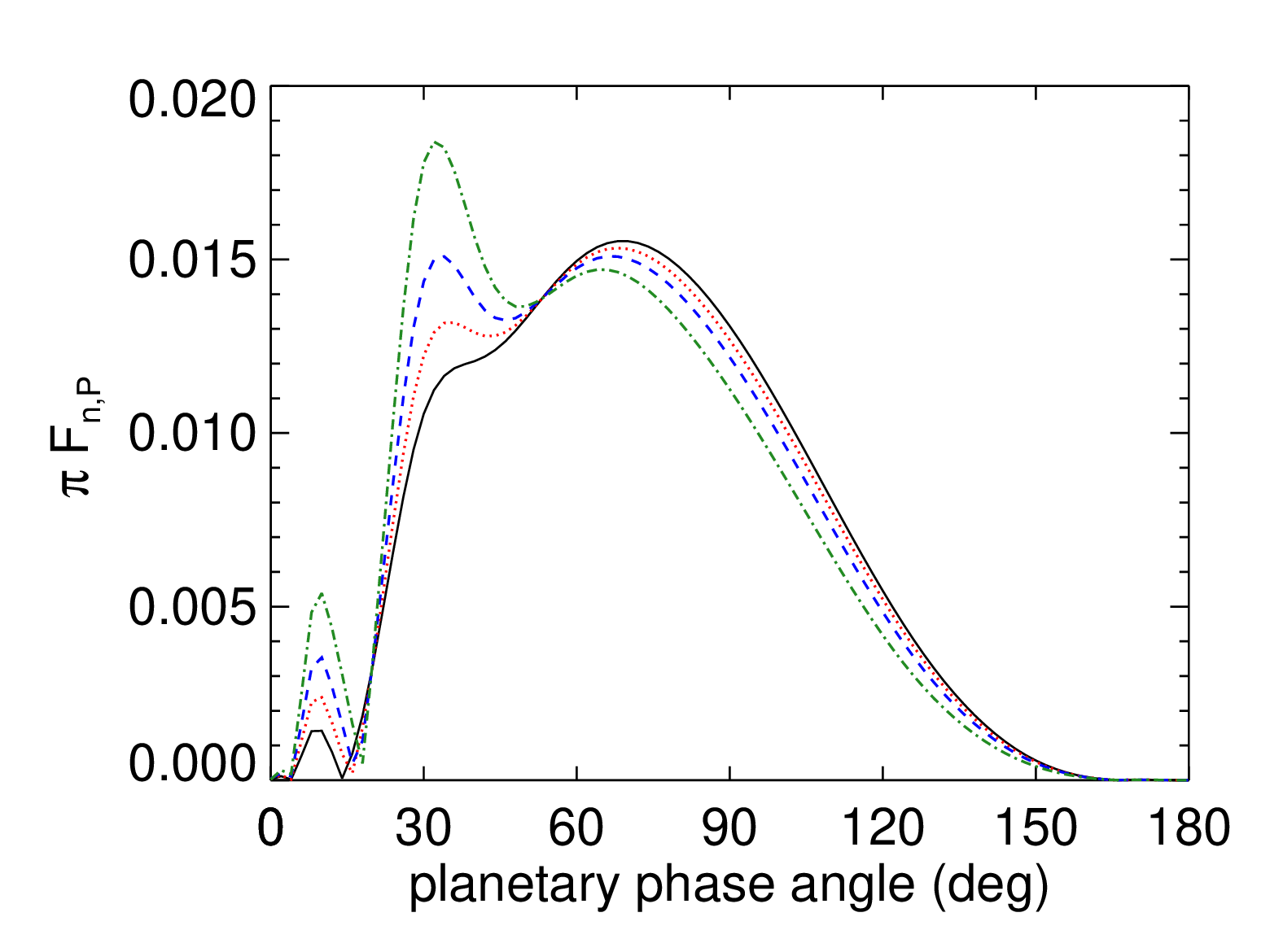}
\hspace{0.8cm}
\centering
\includegraphics[width=85mm]{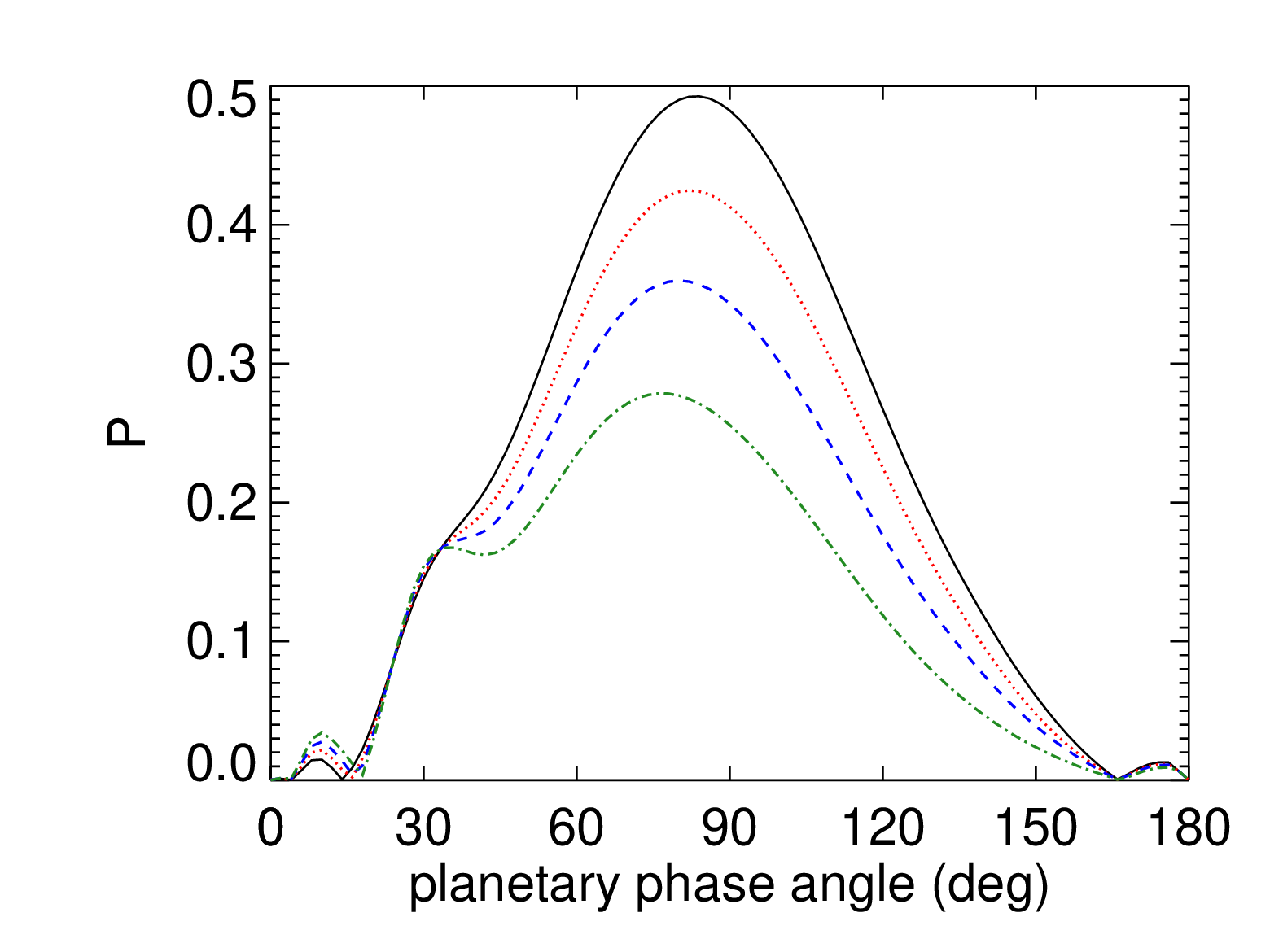}
\caption{Horizontal cuts through the panels of
  Fig.~\ref{fig:all_clcv_fp} for cloud coverages of 20\%, 30\%, 40\%,
  and 60\%.}
\label{fig:all_clcv_fp_cut_through}
\end{figure}

Figure~\ref{fig:all_clcv_fp} shows the total reflected flux $\pi
F_\mathrm{n}$, the polarized reflected flux $\pi F_\mathrm{n,P}$, and
the degree of polarization $P$ at 0.550~$\mu$m, as functions of the
phase angle $\alpha$ and the percentage of cloud coverage. Horizontal
cuts through Fig.~\ref{fig:all_clcv_fp} at a number of coverages are
shown in Fig.~\ref{fig:all_clcv_fp_cut_through}. Because the model
planets are not mirror--symmetric with respect to the planetary
scattering plane, we use Eq.~\ref{eq:poldef} to define the degree of
polarization instead of the $P_\mathrm{s}$ used in the previous
section.

Clearly, as the cloud coverage increases, the flux and polarization
features converge smoothly towards the signals for completely cloudy
planets \citep[see also][for flux and polarization signals of
  completely cloudy planets]{karalidi11}.  At small coverages, $P$ has
a strong maximum around $\alpha=90^\circ$, which is due to scattering
by the gas molecules above and between the patchy clouds.  The
Rayleigh scattering optical thickness above the clouds is 0.06 at
$\lambda=0.550~\mu$m, and between the clouds, it is 0.097 (see
Sect.~\ref{sect_2.3}).
As expected from the single scattering curves
(Fig.~\ref{fig:sing_scat_ice}), the primary rainbow feature is located
close to $\alpha=30^\circ$.  The precise rainbow phase angle depends
on the size of the particles and the wavelength, but, as expected, it
does not depend on the cloud coverage. 

In the total flux $\pi F_\mathrm{n}$, the rainbow is difficult to
discern regardless of the cloud coverage (Fig.~\ref{fig:all_clcv_fp}),
as can also be seen in Fig.~\ref{fig:all_clcv_fp_cut_through}. In the
polarized flux, the rainbow shows up as a local maximum for coverages
of about 30\% or more.  In the degree of polarization, $P$, the
rainbow feature is a shoulder on the Rayleigh scattering maximum from
a cloud coverage of $\sim$20\%. For a cloud coverage larger than about
40\%, the rainbow causes a local maximum in $P$, because the Rayleigh
scattering maximum decreases with increasing cloud coverage.  A
reflecting (i.e. non--black) surface underneath our atmosphere would
increase the contrast of the (primary) rainbow peak by lowering the
intensity of the Rayleigh scattering peak due to the increase of light
with a generally low degree of polarization.

At phase angles near 20$^\circ$, $\pi F_\mathrm{n,P}$ and $P$ are
close to zero, fairly independent of the cloud coverage (see
Fig.~\ref{fig:all_clcv_fp_cut_through}).  This phase angle corresponds
to a neutral point in the single scattering polarization phase
function (Fig.~\ref{fig:sing_scat_ice}). The type~A cloud particle's
neutral point near 76$^\circ$ is lost in the Rayleigh scattering
contribution to $P$. The neutral point near 5$^\circ$ yields the
near--zero polarization region at the smallest phase angles, while the
neutral point near 158$^\circ$ and the generally low polarization
values at those phase angles (see Fig.~\ref{fig:sing_scat_ice}) show
up as the broad low $P$ region at the largest phase angles in
Fig.~\ref{fig:all_clcv_fp}. Note that at these largest phase angles,
the planet is almost in front of its star, and will be extremely
difficult to detect directly anyway.

\begin{figure}
\centering
\includegraphics[width=85mm]{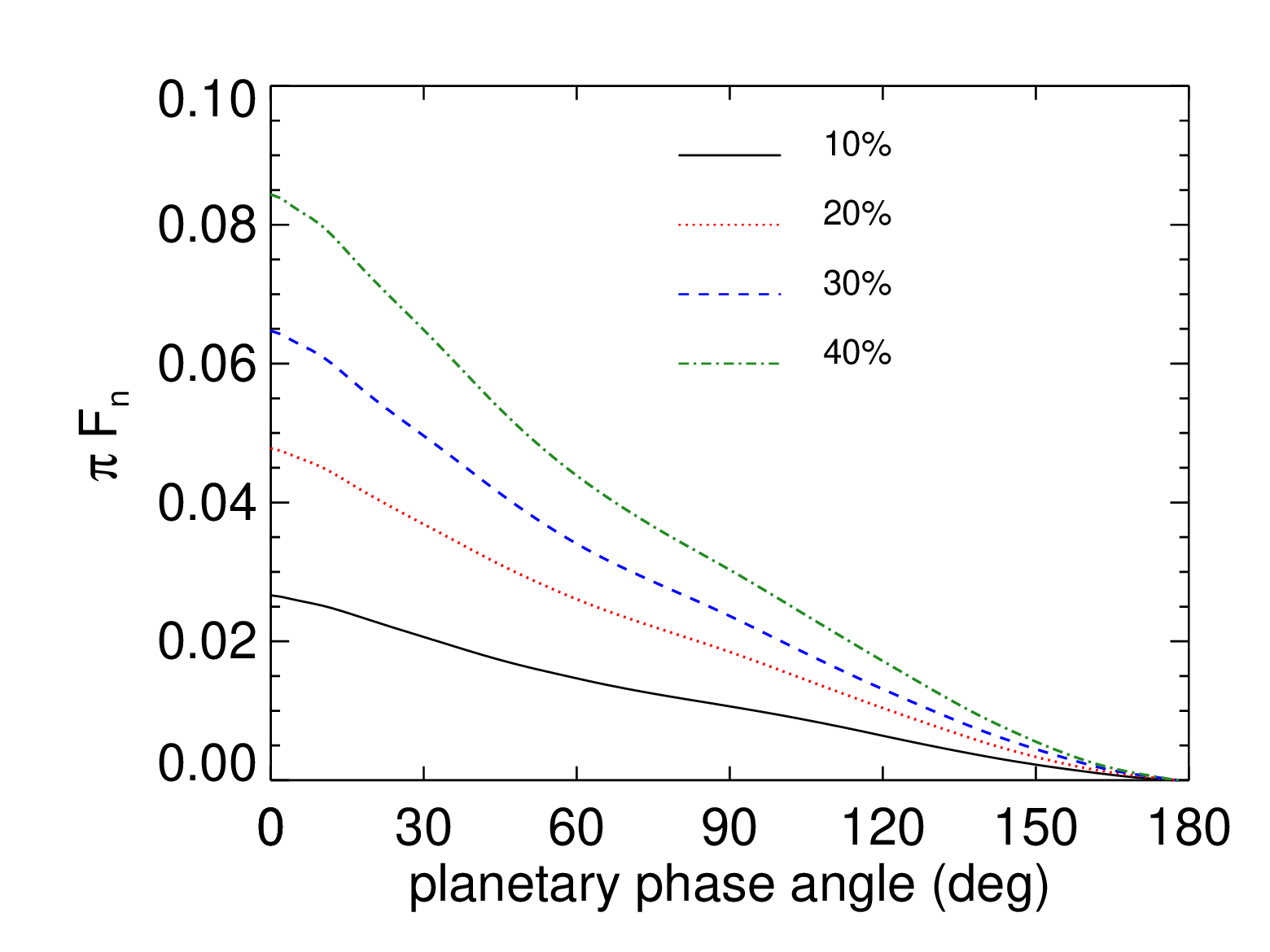}
\hspace{0.8cm}
\centering
\includegraphics[width=85mm]{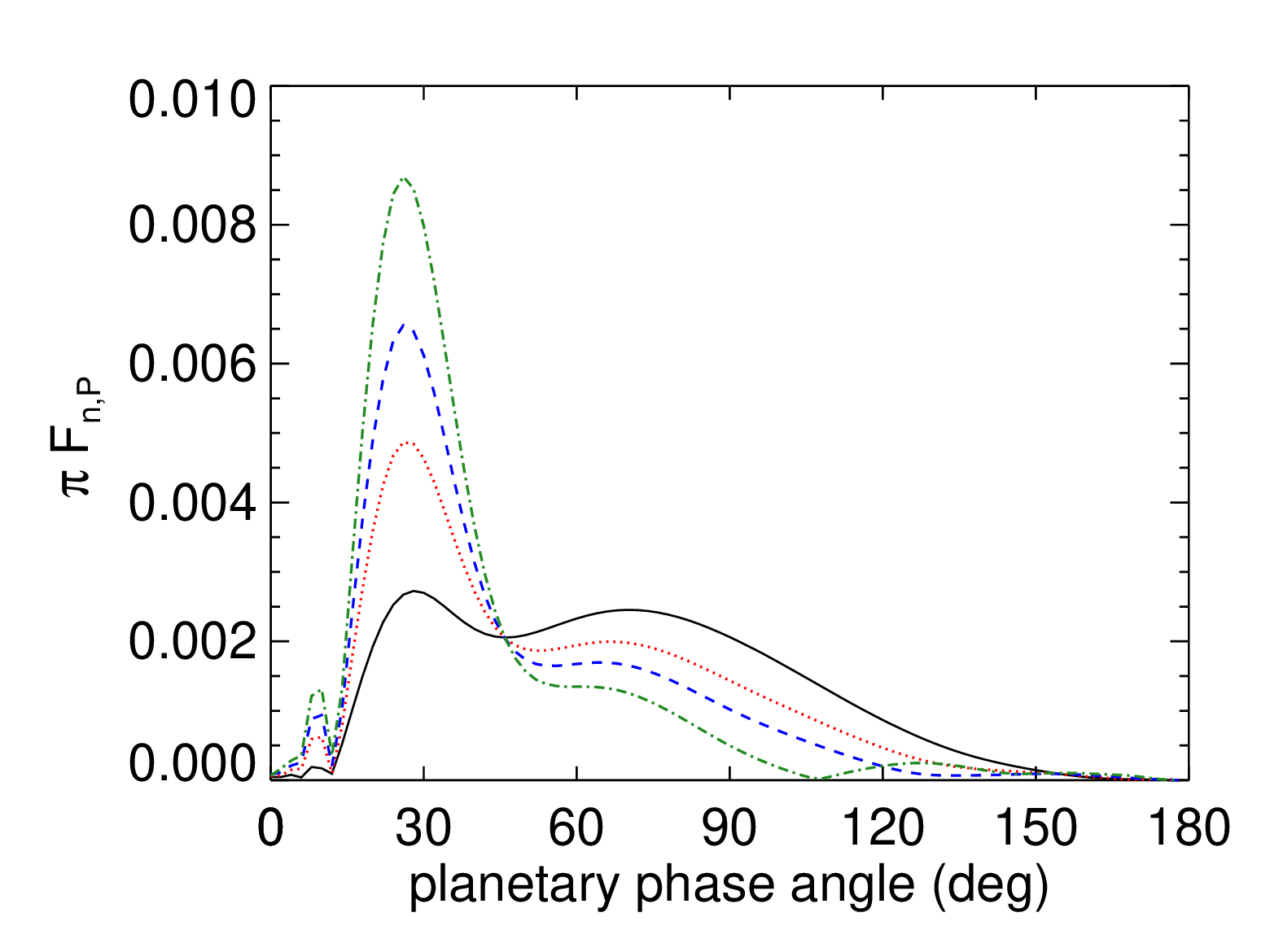}
\hspace{0.8cm}
\centering
\includegraphics[width=85mm]{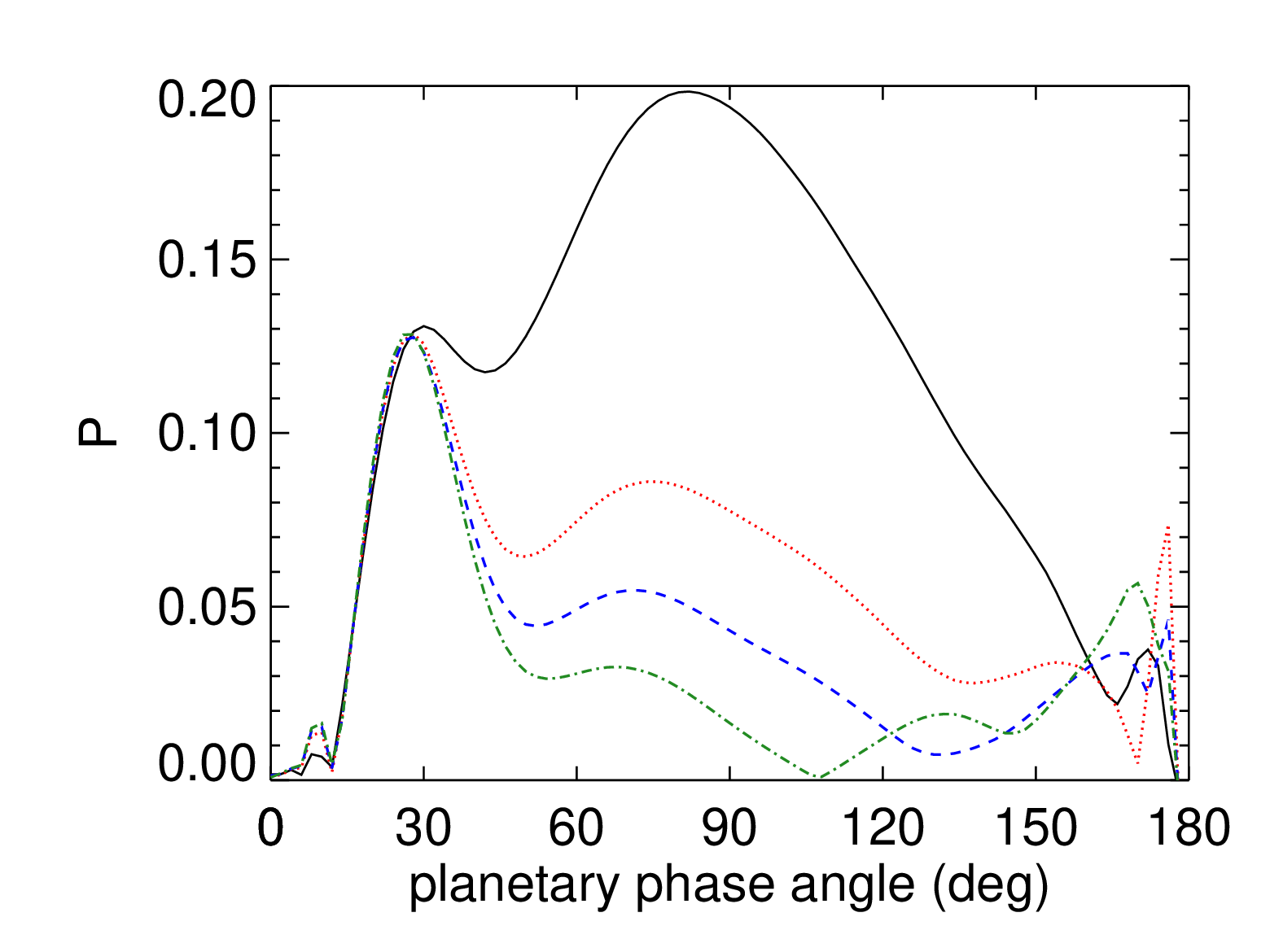}
\caption{Similar to Fig.~\ref{fig:all_clcv_fp_cut_through}, except for
  $\lambda=0.865~\mu$m.}
\label{fig:cut_through_865}
\end{figure}

Figure~\ref{fig:cut_through_865} shows the same as
Fig.~\ref{fig:all_clcv_fp_cut_through}, except for
$\lambda=0.865~\mu$m.  At this wavelength, the Rayleigh scattering
optical thickness above the clouds is 0.01, while between the clouds
it is 0.016 (see Sect.~\ref{sect_2.3}). The optical thickness of the
clouds at 0.865~$\mu$m is 2.1. The differences in reflected fluxes and
the degree of polarization between Fig.~\ref{fig:cut_through_865} and
Fig.~\ref{fig:all_clcv_fp_cut_through} are due to the difference in
Rayleigh optical thickness above and between the clouds and to the
difference in scattering properties of the cloud particles at the
different wavelengths.  Note that the effects of lowering the Rayleigh
scattering optical thickness above the clouds are similar to
increasing the altitude of the clouds.

At 0.865~$\mu$m, the reflected total flux $\pi F_\mathrm{n}$ is a very
smooth function of $\alpha$, almost without even a hint of the
rainbow.  This smoothness is due to the smoother single scattering
phase function of the small type~A particles at this relatively long
wavelength. The degree of polarization shows a very strong rainbow
signature, especially for cloud coverages larger than 10\%.  The
reason that the rainbow is so strong is mostly due to the small
Rayleigh scattering optical thickness above the clouds, which
suppresses the strong polarization maximum around 90$^\circ$.  The
strength of the rainbow appears to be fairly independent of the
percentage of coverage (all clouds have the same optical thickness).
Around $\alpha=108^\circ$, $P$ is virtually zero for a coverage of
40\%.  These low values of $P$ (also in the graphs for the 20\% and
30\% coverage, but then around $\alpha=135^\circ$ and~130$^\circ$,
respectively) are due to the direction of polarization of the light
scattered by the cloud particles (cf. Figs.~\ref{fig:sing_scat_ice}
and~\ref{fig:sing_scat_ice_wavs}) that is opposite to that of light
scattered by the gas molecules.  Because at 0.865~$\mu$m, the Rayleigh
scattering optical thickness is small, the single scattering
polarization signatures of the ice particles dominate the polarization
signature of the planet.

\section{The influence of mixed cloud droplet sizes}
\label{sec:mixed_sizes}

In the previous section, all clouds were composed of the same type of
liquid water cloud droplets. In reality, cloud droplet sizes vary
across the Earth. Typically, droplets above continents are smaller
than those above oceans, due to different types of condensation nuclei
and different amounts of condensation nuclei. In particular, typical
cloud particle radii range from about $5~\mu$m to $30~\mu$m, with a
global mean value of about $8.5~\mu$m above continental areas, and
$11.8~\mu$m above maritime areas \citep{han94}.  Also within clouds,
droplet size distributions will vary, depending e.g. on internal
convective updrafts or downdrafts that influence droplet growth
through condensation and collisions \citep[see
  e.g.][]{stephens87,spinhirne89}.  Here, we will present the
influence of clouds with different liquid water droplet size
distributions on the flux and polarization signals of a model planet.

Our model planet has two layers of clouds, with the lower clouds
located between 1~and 3~km (0.902~and 0.710~bar) and the upper clouds
between 3~and 4~km (0.710~and 0.628~bar)
(cf. Fig.~\ref{fig:profiles}).  The lower clouds are composed of the
type B droplets ($r_\mathrm{eff}=6.0 \mu$m, $v_\mathrm{eff}=0.4$) and
have a coverage of 42.3\%. The upper clouds are composed of the
smaller, type A droplets ($r_\mathrm{eff}=2.0 \mu$m,
$v_\mathrm{eff}=0.1$) and have a coverage of 16\%. The maps for the
upper and the lower clouds were generated separately and hence there
are regions with only lower clouds, or only upper clouds, or both.

\begin{figure}
\centering 
\includegraphics[width=85mm]{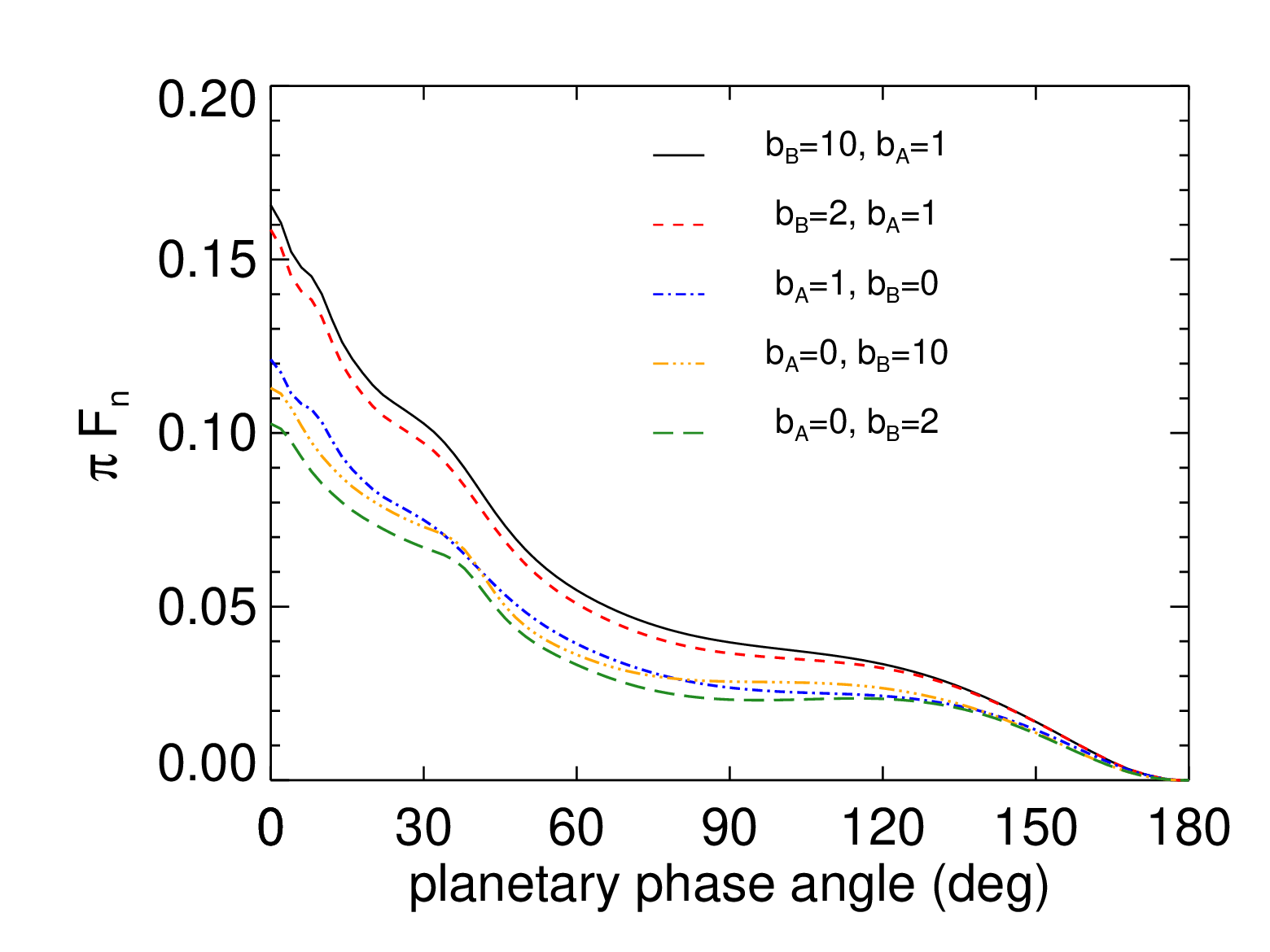}
\hspace{0.8cm}
\centering 
\includegraphics[width=85mm]{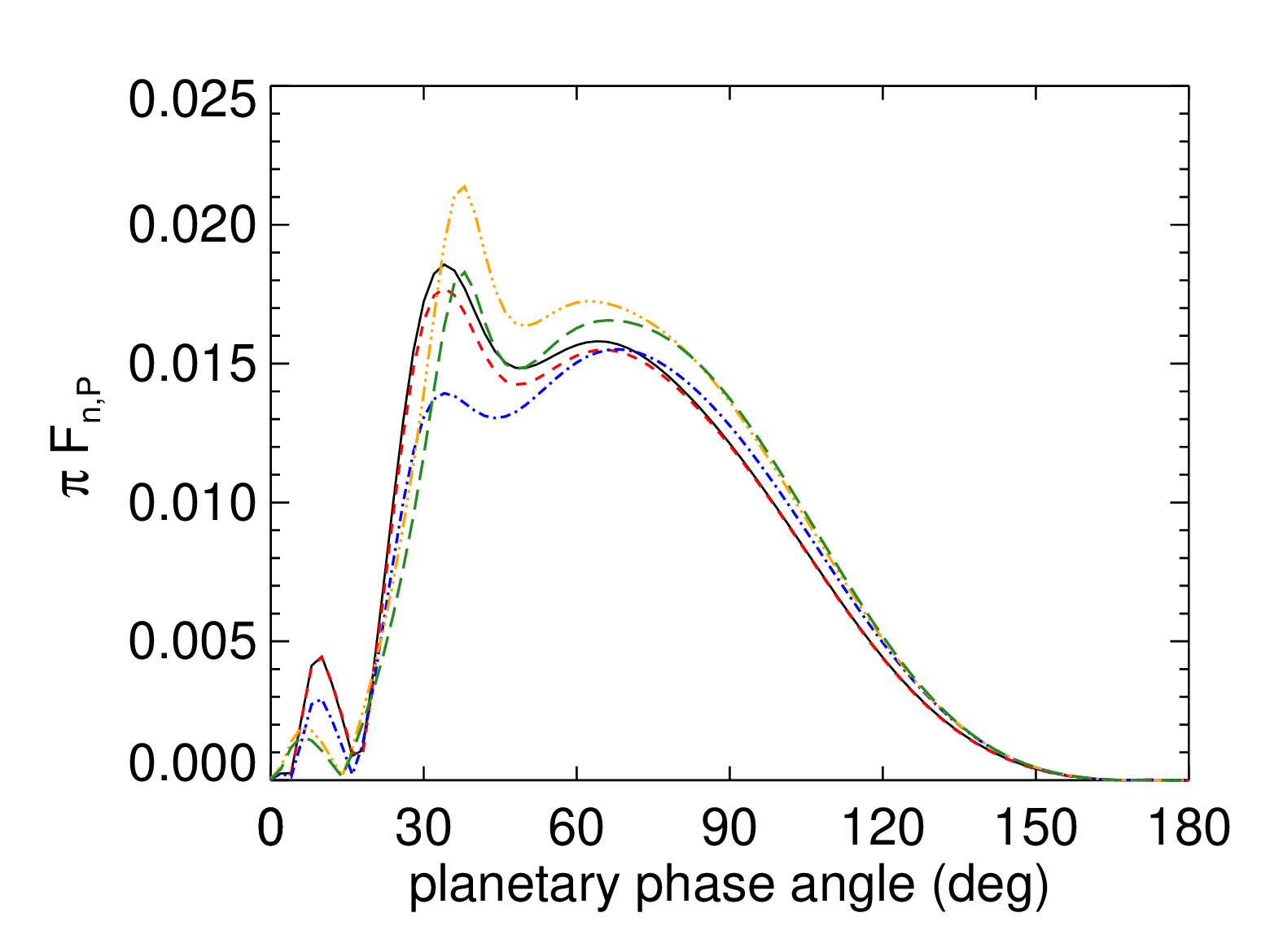}
\hspace{0.8cm}
\centering
\includegraphics[width=85mm]{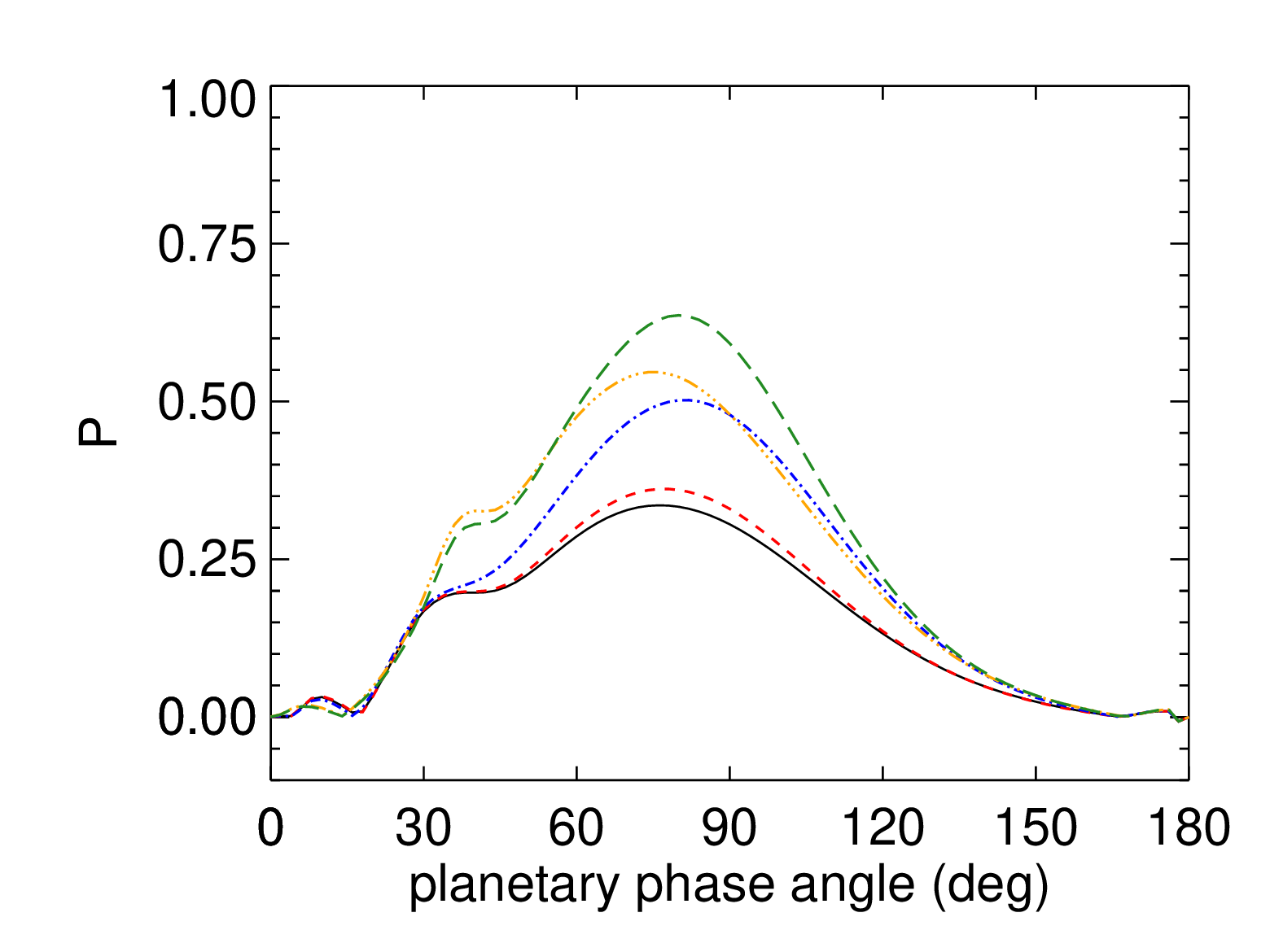}
\caption{$\pi F_\mathrm{n}$, $\pi F_\mathrm{n,P}$, and $P$ as
  functions of $\alpha$ for a planet with a lower, type~B liquid water
  cloud that covers $\sim$42.3\% and an upper, type~A liquid water
  cloud layer that covers $\sim$16\% of the planet for various values
  of the cloud optical thicknesses.}
\label{fig:dbllyr_AB}
\end{figure}

Figure~\ref{fig:dbllyr_AB} shows the reflected total flux $\pi
F_\mathrm{n}$, the polarized flux $\pi F_\mathrm{n,P}$, and the degree
of polarization $P$ for our model planet at $\lambda=0.550~\mu$m for
different values of the cloud optical thickness, $b$. Because the two
types of cloud particles have slightly different locations of the
rainbows, the rainbow in the total flux in the presence of two cloud
layers appears to be somewhat broadened, as compared to the curves for
the fluxes of single layers of clouds (the curves in
Fig.~\ref{fig:dbllyr_AB} where one of the cloud optical thicknesses
equals zero).

In $P$, the strength of the rainbow feature appears to depend on the
properties of the highest cloud layer: adding a lower cloud layer
mainly decreases the strength of the Rayleigh scattering maximum by
adding more low polarized light at these phase angles.  $P$ is fairly
insensitive to the optical thickness of this lower cloud layer (i.e. 2
or 10): increasing the thickness of this cloud slightly increases $\pi
F_\mathrm{n}$, but also slightly increases $\pi F_\mathrm{n,P}$.  This
latter effect can also be seen in the rainbow feature in the curves
for a planet with a single, lower cloud layer: with increasing $b$,
$\pi F_\mathrm{n}$ increases, but $\pi F_\mathrm{n,P}$ increases, too!
At larger phase angles ($\alpha > 55^\circ$ in
Fig.~\ref{fig:dbllyr_AB}), increasing $b$ of the lower cloud layer
increases $\pi F_\mathrm{n}$ but does not increase $\pi
F_\mathrm{n,P}$ significantly, which can be attributed to the single
scattering properties of the cloud droplets (see
Fig.~\ref{fig:sing_scat_ice}).

\section{The influence of the ice cloud coverage} 
\label{sec:h2oice}

On Earth, liquid water clouds can be overlaid by water ice clouds.
These latter clouds can be entirely separate (such as high-altitude
cirrus clouds) or the upper parts of vertically extended clouds that
have lower parts consisting of liquid water droplets.  The amount of
ice cloud coverage on Earth depends strongly on the latitude. Cirrus
clouds for example, have an annual mean coverage of $\sim$20\% in the
tropics, while at northern mid-latitudes the annual mean coverage goes
down to $\sim$12\%. Globally, cirrus clouds cover $\sim$14\% of the
Earth's upper troposphere \citep[][]{eleftheratos07}.  As mentioned in
Sect.~\ref{sect:intro}, water ice particles come in various sizes and
shapes that yield single scattering properties that differ from those
of liquid water droplets. In particular, light that has been singly
scattered by ice cloud particles does not show a rainbow feature
between phase angles of 30$^\circ$ and 40$^\circ$ (i.e. between single
scattering angles of 140$^\circ$ and 150$^\circ$).

To investigate the influence of water ice clouds on the strength of
the rainbow feature of liquid water clouds, we used model planets with
two layers of clouds: a lower layer of optical thickness 10,
consisting of type~B liquid water droplets and an upper layer
containing water ice particles \citet[][]{hess94, hess98}. The lower
layer is located between 279~and 273~K (3~and 4~km, or 0.710~and
0.628~bar) and the upper layer between 253~and 248~K (7~and 8~km, or
0.426 and 0.372~bar) (c.f. Fig.~\ref{fig:profiles}).  The single
scattering properties of the water ice particles clouds have been
calculated using the code of \citet[][]{hess94, hess98}. The singly
scattered total and polarized flux and the degree of polarization of
our ice cloud particles are shown in Fig.~\ref{fig:sing_scat_ice} for
$\lambda=0.550~\mu$m and in Fig.~\ref{fig:sing_scat_ice_wavs} for
$\lambda=0.660$ and 0.865$~\mu$m.

\begin{figure}
\centering \includegraphics[width=85mm]{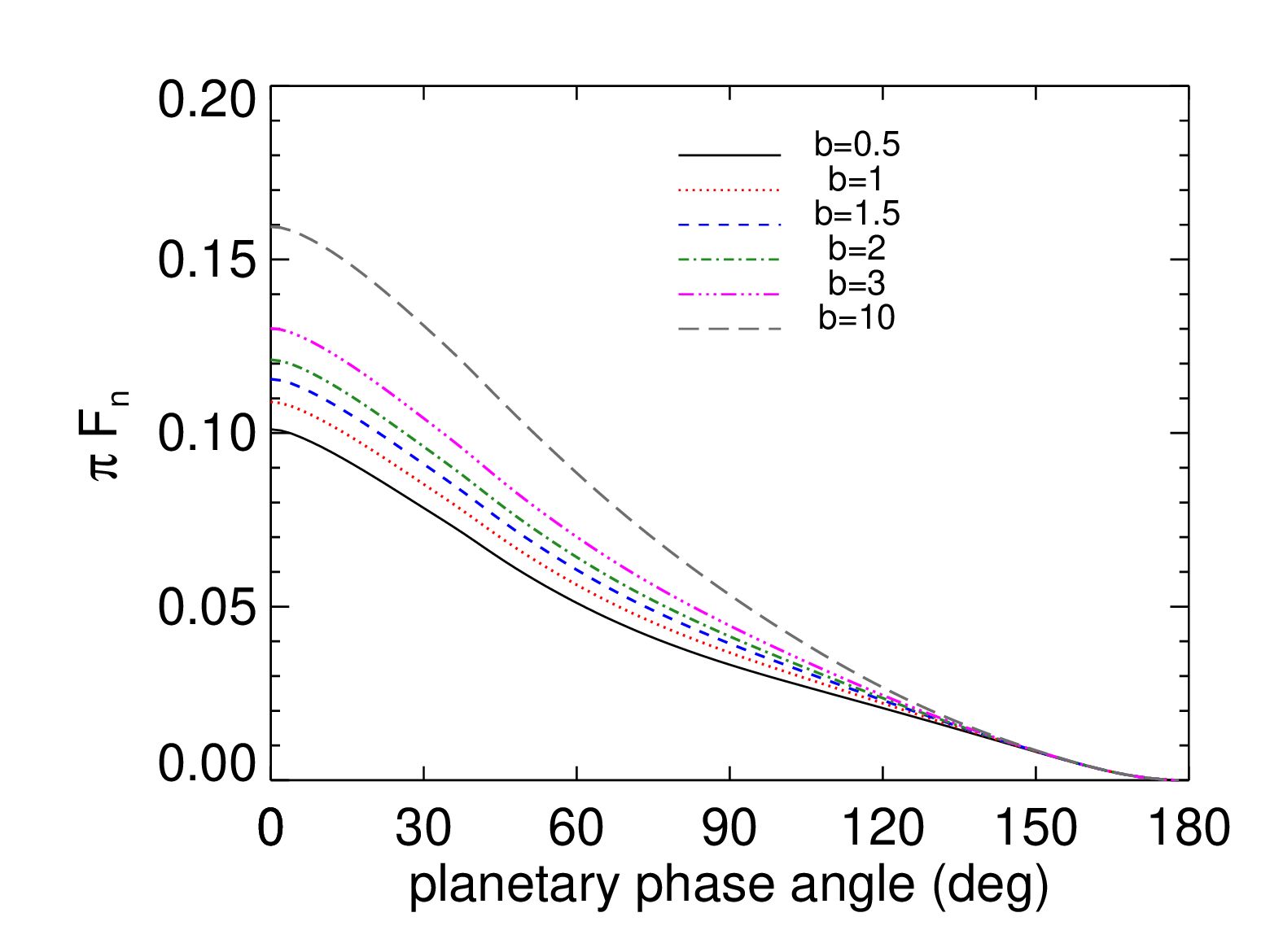}
\hspace{0.8cm}
\centering
\includegraphics[width=85mm]{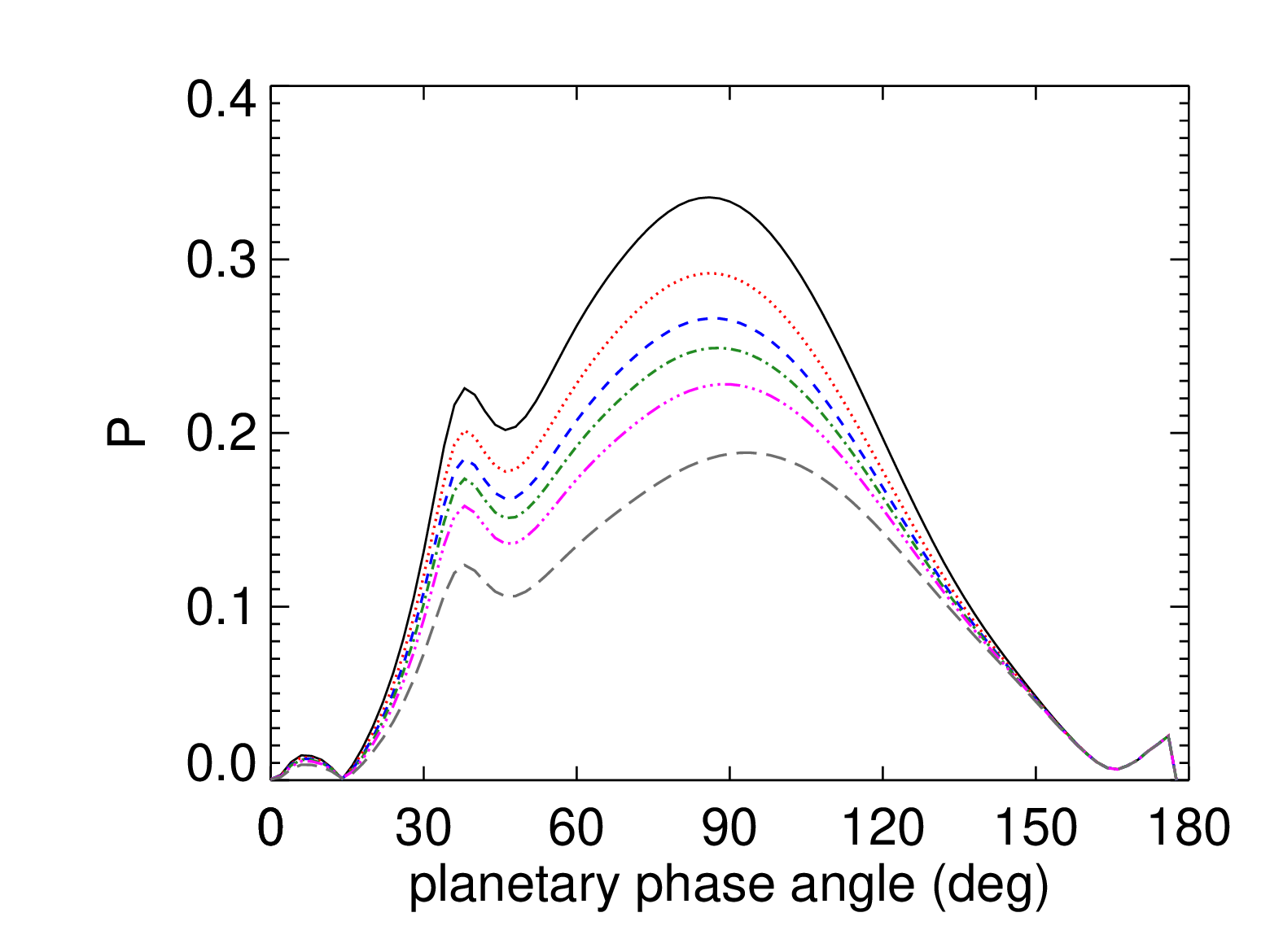}
\caption{$\pi F_\mathrm{n}$ and $P$ as functions of $\alpha$ at
  $\lambda=0.550~\mu$m, for a planet with a lower liquid water cloud
  with $b=10$ that covers $\sim$42.3\% of the planet and an upper ice
  cloud layer that covers $\sim$16\% of the planet, and $\sim$11\% of
  the liquid water clouds.}
\label{fig:icevar_b}
\end{figure}

Our first model planets have a liquid water cloud layer that covers
$\sim$~42.3\% of the surface, and an ice cloud layer that covers
$\sim$~16\%. About 4.5\% of the liquid water clouds is covered by ice
clouds (i.e. $\sim$~12\% of the ice clouds is covering a liquid water
cloud).  Figure~\ref{fig:icevar_b} shows $\pi F_\mathrm{n}$ and $P$ of
our model planets as functions of $\alpha$ for various optical
thicknesses of the ice clouds.

Adding a thin ice cloud (with $b=0.5$) to the cloudy planet lowers
$\pi F$ at the smallest phase angles (cf. Fig.~\ref{fig:dbllyr_AB}),
because the ice particles are less less strong backward scattering
than the liquid water droplets (cf. Fig.~\ref{fig:sing_scat_ice}). At
intermediate phase angles (around $\alpha=90^\circ$), the thin ice
clouds brighten the planet, because at those angles, their single
scattering phase function is higher than that of the liquid water
droplets (cf. Fig.~\ref{fig:sing_scat_ice}).  Clearly, with increasing
ice cloud optical thickness the reflected flux increases across the
whole phase angle range (which is difficult to see at phase angles
larger than about 150$^\circ$).

The total flux $\pi F_\mathrm{n}$ does not show any evidence of a
rainbow feature, not even for the thinnest ice cloud.  The degree of
polarization $P$, however, clearly shows the signature of the rainbow,
even for large optical thicknesses of the ice cloud particles
(cf. Fig.~\ref{fig:icevar_b}). The rainbow shows up in $P$ despite the
overlaying ice clouds, because the ice particles themselves have a
very low polarization signature especially in the phase angle region
of the rainbow (see Fig.~\ref{fig:sing_scat_ice}), so the only
polarized signals arise from scattering by the liquid cloud particles
and the gas molecules (which yields the maximum in $P$ around
90$^\circ$).  Increasing the ice cloud optical thickness, decreases
$P$ across all phase angles because it increases the amount of mostly
unpolarized reflected light.

\begin{figure}
\centering
\includegraphics[width=85mm]{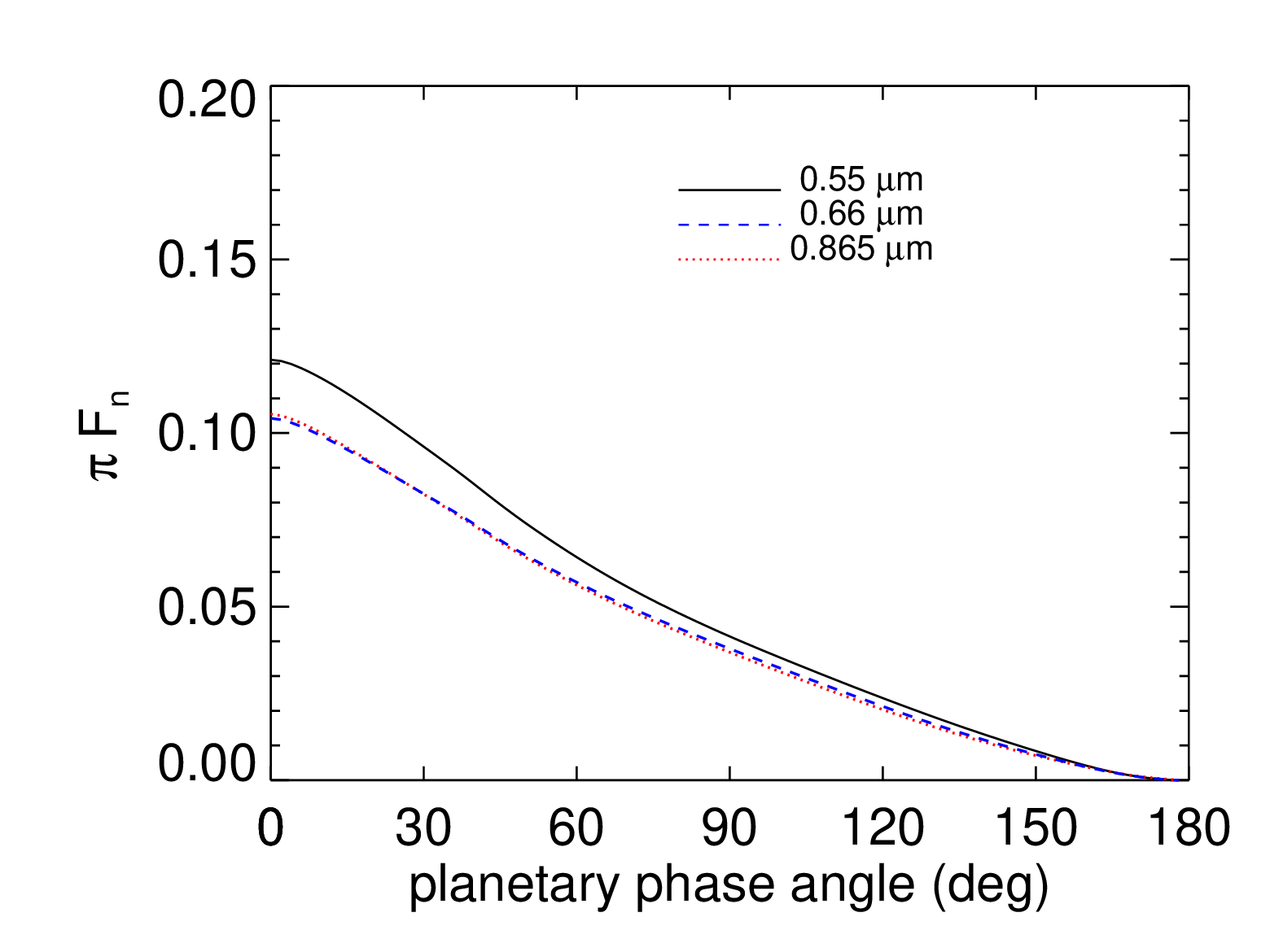}
\hspace{0.8cm}
\centering
\includegraphics[width=85mm]{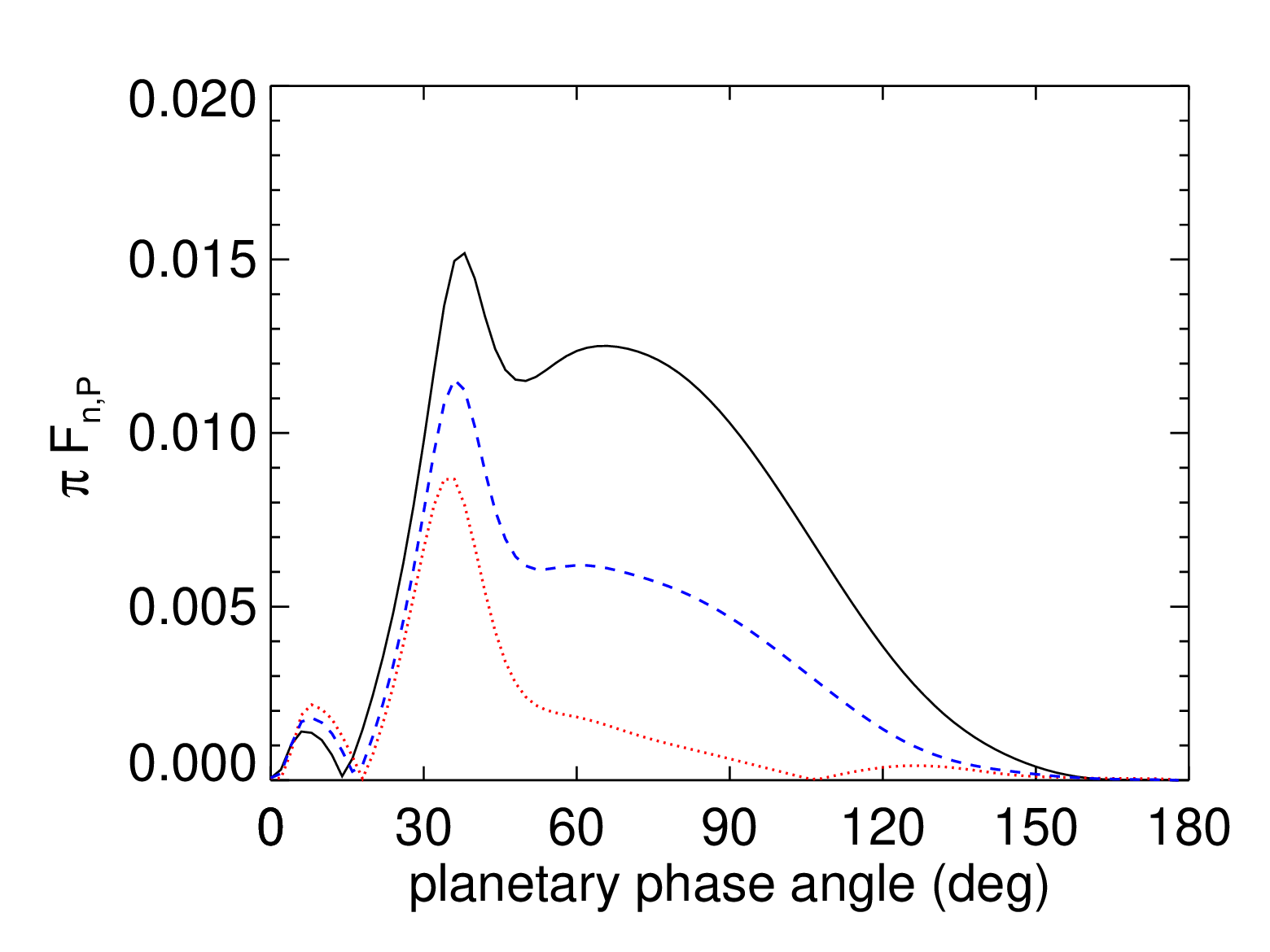}
\hspace{0.8cm}
\centering
\includegraphics[width=85mm]{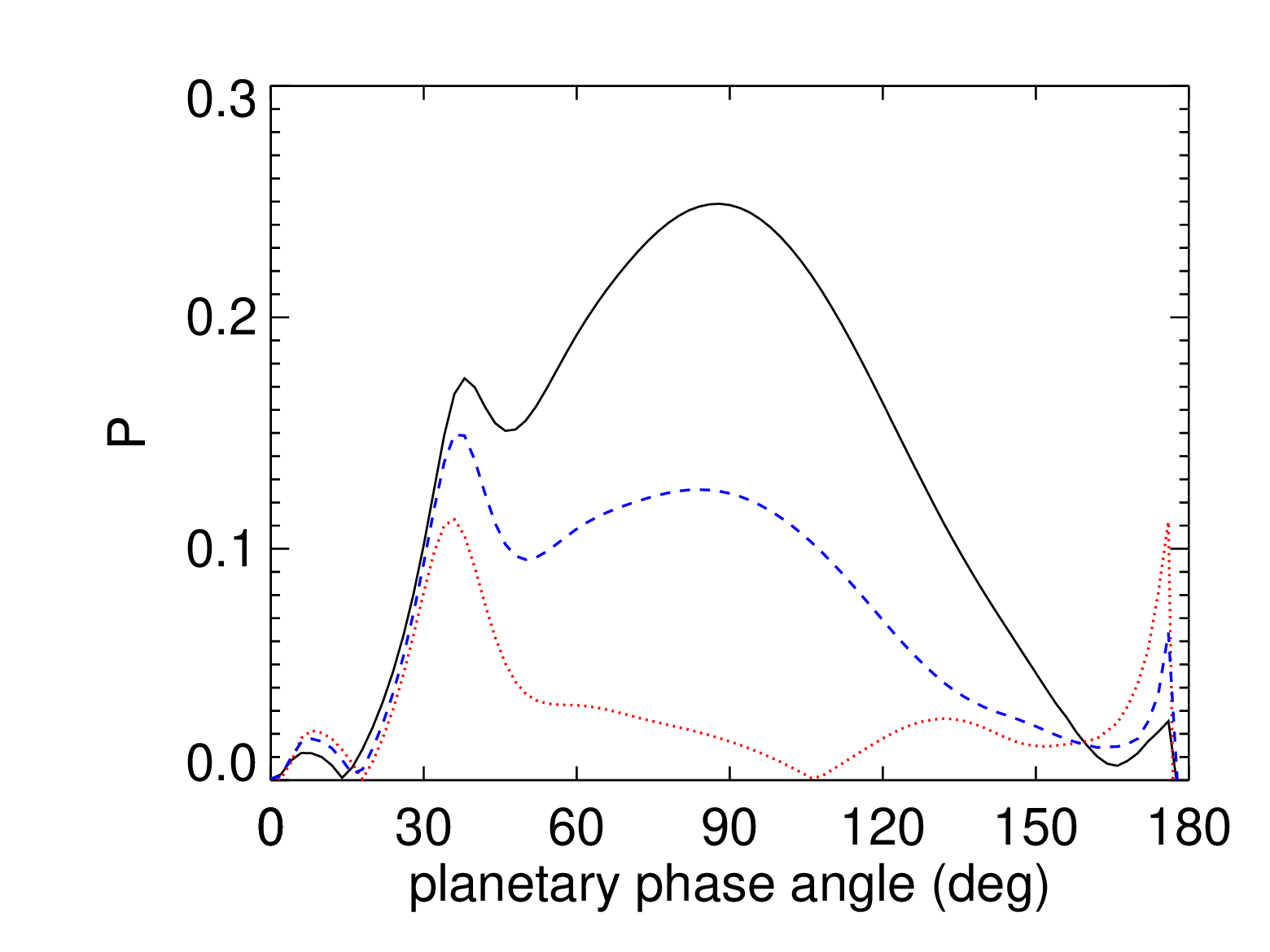}
\caption{Similar to Fig.~\ref{fig:icevar_b}, except for an ice cloud
  with $b=2.0$, and for $\lambda=0.550$, 0.660, and 0.865~$\mu$m. The
  $\pi F_{\rm n}$ curves for 0.660 and 0.865~$\mu$m overlap.}
\label{fig:wavs_cloud_4216}
\end{figure}

In Fig.~\ref{fig:wavs_cloud_4216}, we have plotted $\pi F_\mathrm{n}$,
$\pi F_\mathrm{n,P}$ and $P$ for the same model planet as used for
Fig.~\ref{fig:icevar_b}, except for an ice cloud optical thickness of
2.0, and for $\lambda=0.660$~and 0.865~$\mu$m. For comparison, the
$\lambda=0.550$~$\mu$m curves (see Fig.~\ref{fig:icevar_b}) are also
shown. With increasing wavelength, the total reflected flux decreases,
mainly because the Rayleigh scattering optical thickness above and
between the clouds decreases. The decrease of the Rayleigh scattering
optical thickness is also apparent from the decrease of the maximum in
$P$ around 90$^\circ$, and the stronger influence of the single
scattering polarization phase function of the liquid water cloud
droplets. The ice particles leave no obvious feature in the
polarization phase function at longer wavelengths, except that they
decrease $P$ somewhat as compared to $P$ of a planet with only a
liquid cloud (cf. the 40\% coverage curve in
Fig.~\ref{fig:cut_through_865}).

\begin{figure}
\centering \includegraphics[width=85mm]{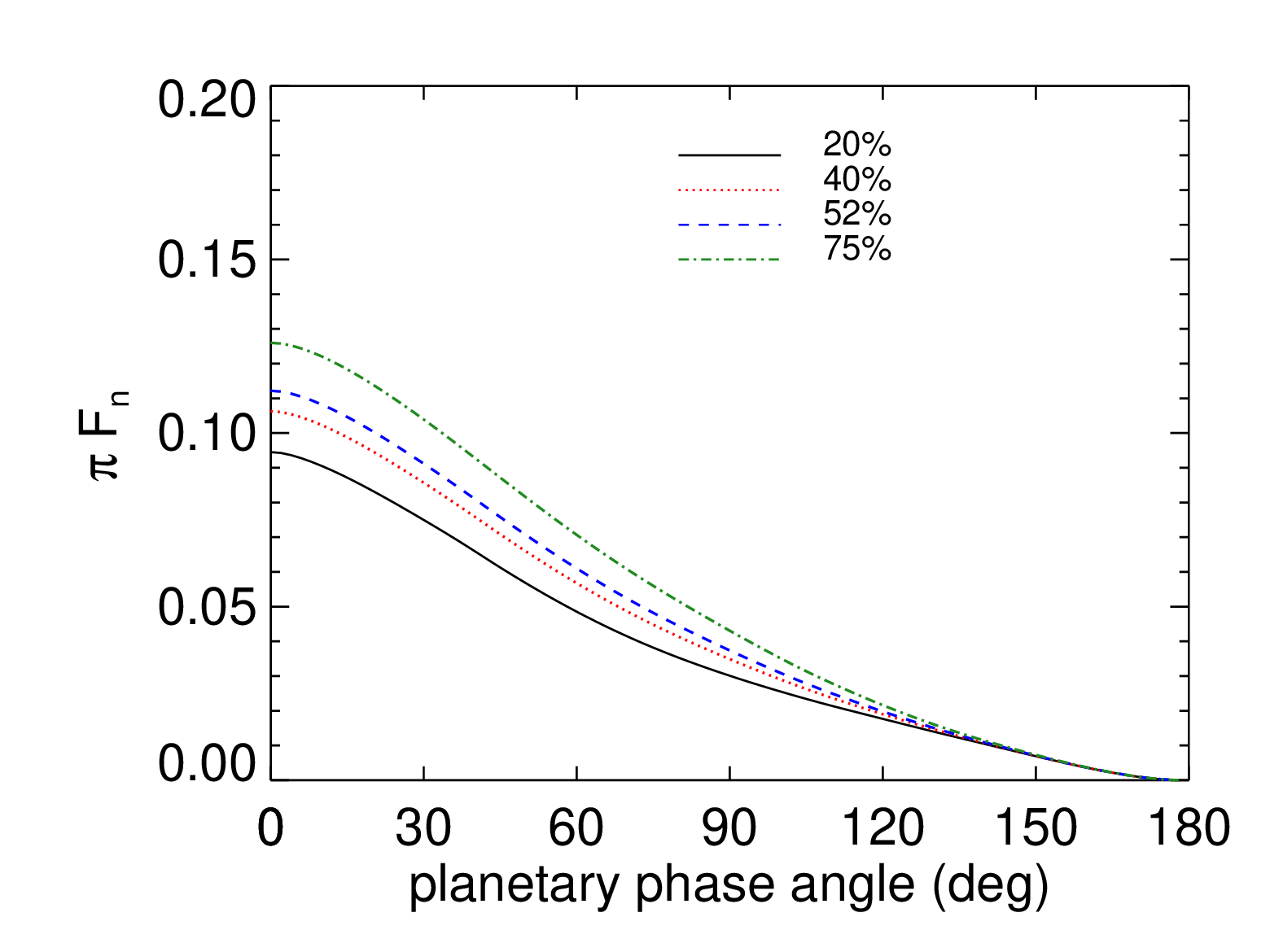}
\hspace{0.8cm}
\centering
\includegraphics[width=85mm]{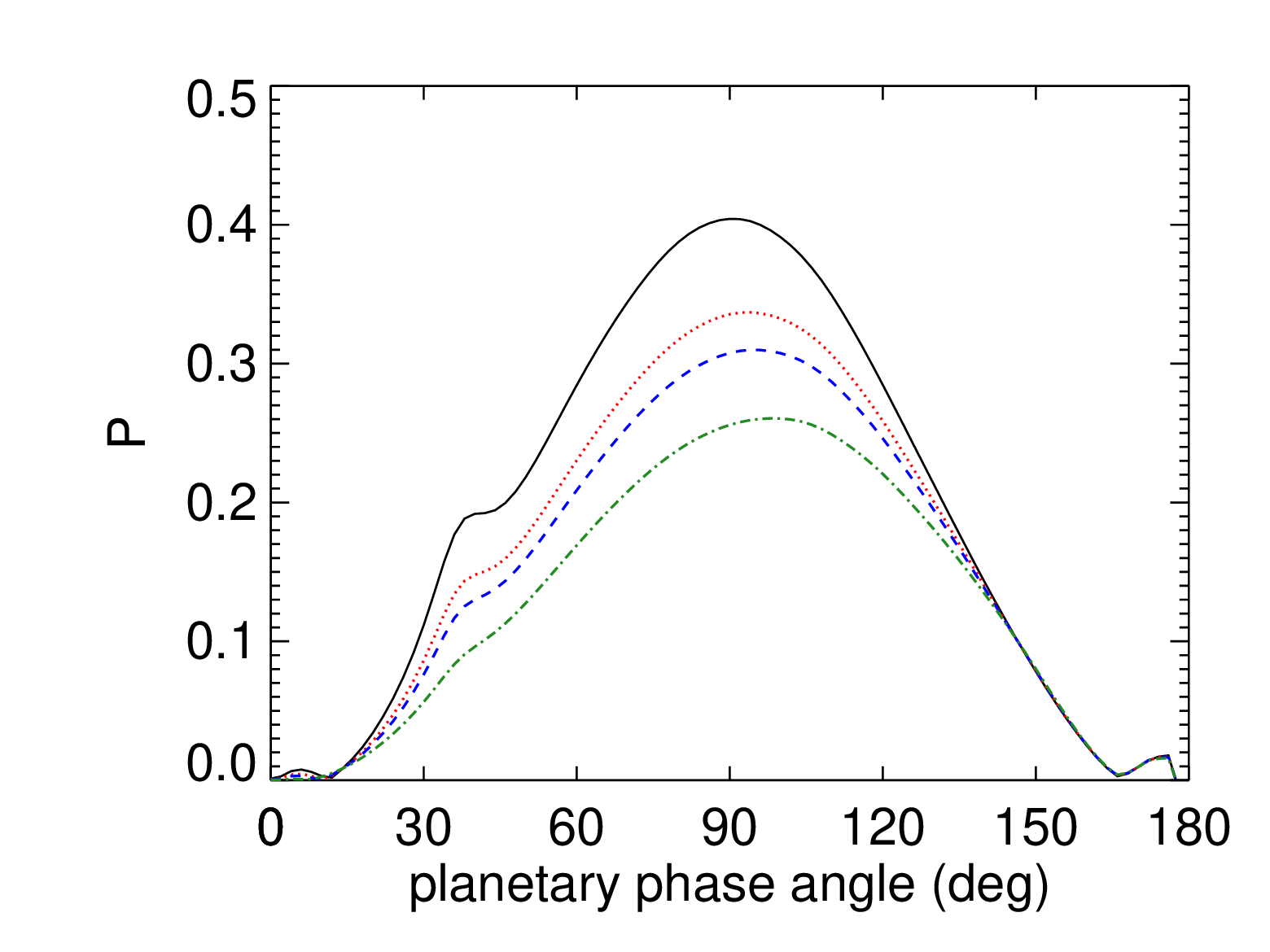}
\caption{Similar to Fig.~\ref{fig:icevar_b}, except for a lower liquid
  water cloud that covers $\sim$25\% of the planet and an upper ice
  cloud with $b=4.0$. The ice cloud coverage is varied from 20\% to
  75\% of the water clouds.}
\label{fig:dbllr_h2o_ice_var_covs}
\end{figure}

To investigate the influence of the ice cloud coverage over liquid
water clouds on the reflected flux and polarization, we used a model
planet with lower liquid water clouds with $b=10$ and a coverage of
$\sim~25$\%, and upper ice clouds with $b=4$ right above the liquid
water clouds. Figure~\ref{fig:dbllr_h2o_ice_var_covs} shows the
planet's flux and polarization phase functions at $\lambda=0.550~\mu$m
for various ice cloud coverages of the liquid water clouds (thus, an
ice cloud coverage of 75\% of the liquid water clouds, equals an ice
cloud coverage of $\sim$16\% of the planet).

\begin{figure}
\centering
\includegraphics[width=85mm]{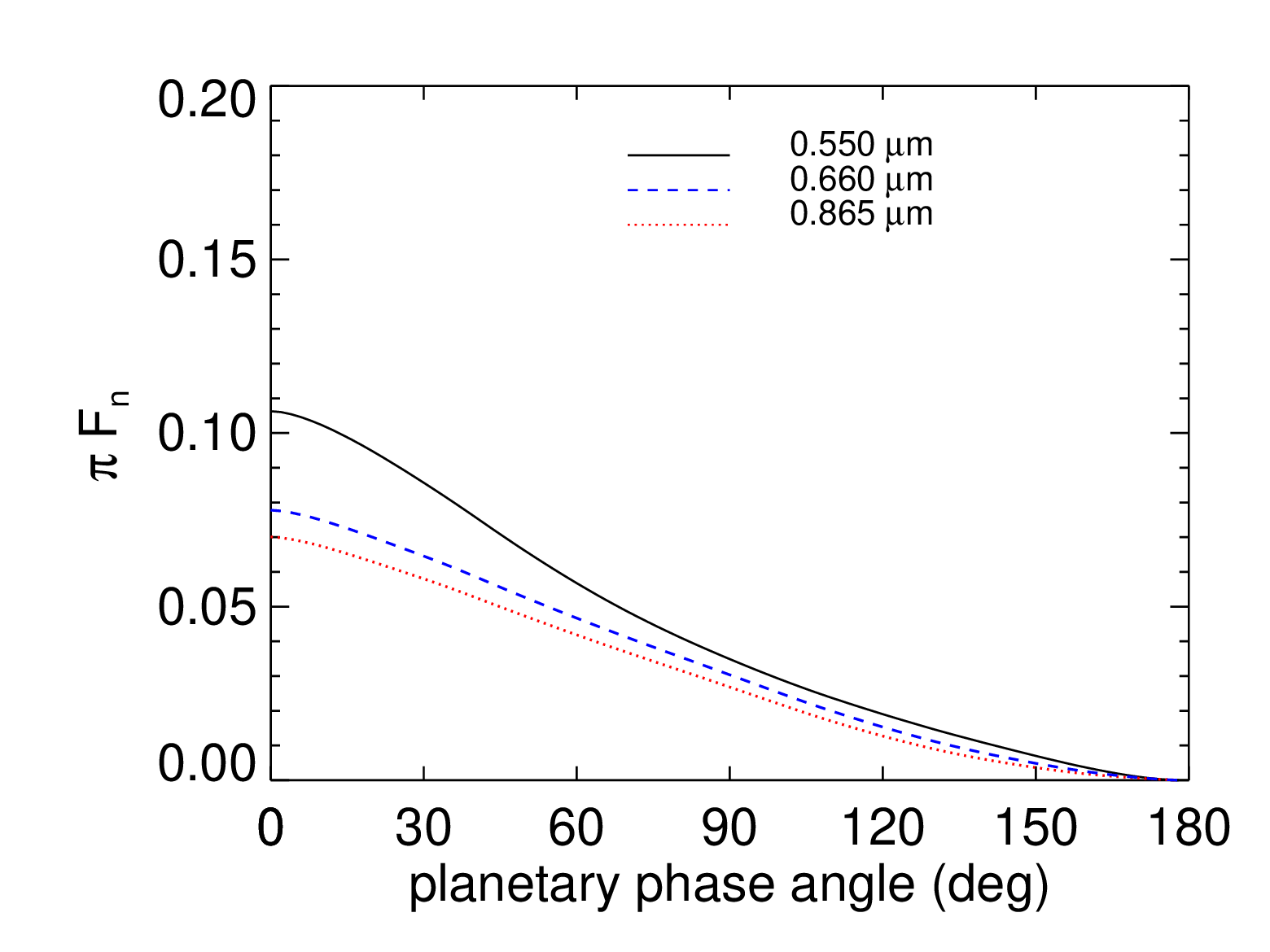}
\hspace{0.8cm}
\centering
\includegraphics[width=85mm]{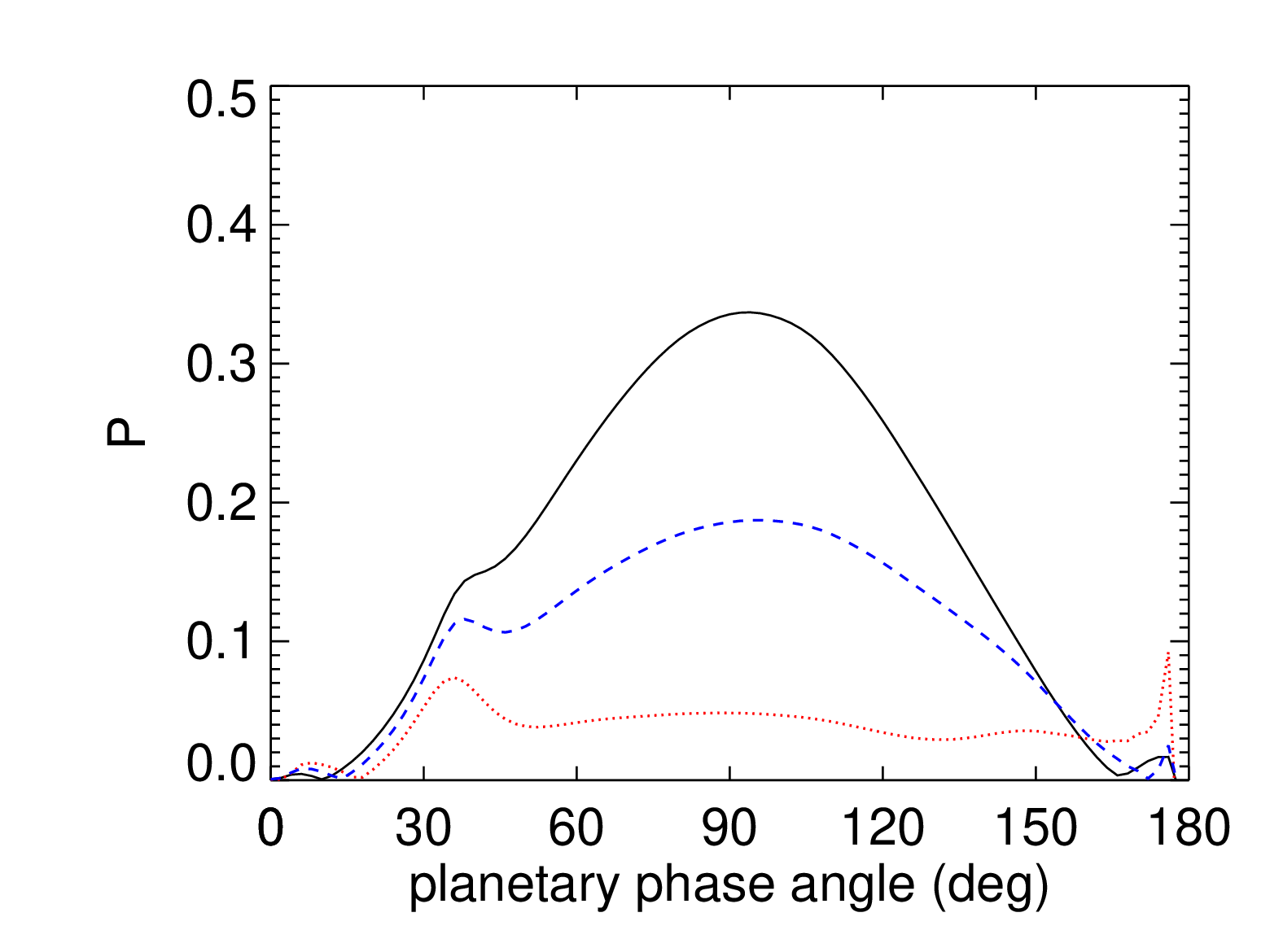}
\caption{Similar to Fig.~\ref{fig:dbllr_h2o_ice_var_covs}, for ice
  clouds covering 40\% of the liquid water clouds, and for
  $\lambda=0.550$, 0.660, and 0.865~$\mu$m.}
\label{fig:dbllr_h2o_ice_var_lmds}
\end{figure}

Even with ice clouds covering as little as 20\% of the liquid water
clouds (5\% of the planet, but only above the liquid water clouds),
the total flux phase function fails to show a hint of the
rainbow. This was also apparent from Fig.~\ref{fig:icevar_b}, where
11\% of the liquid water clouds was covered by an ice cloud.  The
polarization phase function does show the rainbow feature, but up from
an ice cloud coverage of 40\% (a coverage of 10\% of the planet) in
Fig.~\ref{fig:icevar_b}, the feature is increasingly weak such that it
disappears into the Rayleigh scattering polarization maximum around
90$^\circ$.  At longer wavelengths, such as 0.865~$\mu$m, where the
Rayleigh scattering optical thickness above and between the clouds is
smaller, the rainbow feature is still clearly visible in $P$ when ice
clouds cover 40\% of the liquid water clouds. This can be seen in
Fig.~\ref{fig:dbllr_h2o_ice_var_lmds}.

\section{Looking for the rainbow on Earth}
\label{sect_signs_earth}

Figure~\ref{fig:dbllr_h2o_ice_var_covs} shows us that liquid water
clouds on a planet can be detected using polarization measurements of
the reflected starlight, even when the liquid water clouds are
(partly) covered by water ice clouds.  The question arises whether the
rainbow would be visible for a distant observer of the Earth. As far
as we know, the only polarization observations of the Earth have been
done by the POLDER instrument (POLarization and Directionality of the
Earth's Reflectances) a version of which is currently flying onboard
the PARASOL satellite \citep[for a description of the instrument,
  see][]{deschamps94}. Locally, POLDER has indeed observed the primary
rainbow above liquid water clouds \citep[see][]{goloub00}. However,
since PARASOL is in a low--Earth--orbit, the POLDER measurements are
not representative for polarization observations of the whole
(disk--integrated) Earth observed from afar.
\begin{table}[t]
\caption{The cloud optical thicknesses $b$ (at $\lambda= 0.550~\mu$m) 
         from the MODIS/Aqua data from April 25, 2011.
         To avoid having to include too many different values of $b$,
         we have binned the optical thicknesses as shown below. Ice clouds 
         with $b \geq 20$ are ignored.}
\centering
\begin{tabular}{cc|cc}
\multicolumn{2}{c}{Liquid water cloud} & 
\multicolumn{2}{c}{Ice water cloud} \\
\hline \hline
\textbf{$b_\mathrm{MODIS}$} & \textbf{$b_\mathrm{model}$} &
\textbf{$b_\mathrm{MODIS}$} & \textbf{$b_\mathrm{model}$} \\
\hline \hline
$ 0 <    b <  1$ & 0.5 &  $0 < b < 1$ & 0.5 \\
$ 1 \leq b < 10$ & 5   &  $1 \leq b < 2$ & 1.5 \\
$ 5 \leq b < 20$ & 15  &  $2 \leq b < 5$ & 3 \\
$20 \leq b < 50$ & 35  &  $5 \leq b < 10$ & 7.5 \\
$50 \leq b < 80$ & 65  & $10 \leq b < 20$ & 15 \\
\end{tabular}
\label{table:modis_model}
\end{table}

In the absence of real polarization observations, we have simulated
the total flux and polarization of light reflected by the whole Earth
using cloud properties derived from observations by MODIS (Moderate
Resolution Imaging Spectroradiometer), onboard NASA's Aqua
satellite. We used MODIS' cloud coverage (the horizontal distribution
of the clouds), cloud thermodynamic phase (liquid or ice), and the
cloud optical thickness as measured on April 25th, 2011, to build a
model Earth. To limit the number of (time--consuming) calculations, we
binned the measured optical thicknesses according to
Table~\ref{table:modis_model}. Ice clouds with an optical thicknesses
larger than 20 (at $\lambda=0.550~\mu$m) were ignored, so as to
include only cirrus/cirrostratus ice clouds in our sample (according
to the ISCCP categorization) and to avoid deep convection clouds. We
assume that the liquid water clouds consist of the type~B droplets and
for the ice particles, we use the models of \citet{hess94} and
\citet{hess98}. Finally, our ice clouds are positioned at altitudes
with temperatures lower than 253~K, so that we avoid mixed--phase
clouds (in which droplets and crystals exist side--by--side).  The
surface is assumed to be black as an ocean.

\begin{figure}
\centering
\includegraphics[width=85mm]{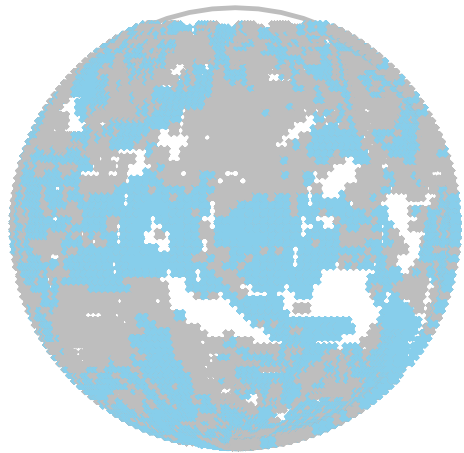}
\caption{The cloud map of the Earth on April 25th, 2011, based on
  MODIS/Aqua data. The planet is covered by $\sim$63\% liquid water
  clouds (gray regions) and $\sim$36\% ice clouds (blue
  regions). About 28\% of the planet is covered by two layers of
  clouds.}
\label{fig:modis_map}
\end{figure}

Figure~\ref{fig:modis_map} shows the cloud map of our model Earth.  On
April 25th, 2011, about 85\% of the planet was covered by clouds
(liquid and/or ice), about 14\% was covered by both liquid and ice
clouds, about 63\% of the planet was covered by liquid water clouds
and about 36\% by water ice clouds.

\begin{figure}
\centering 
\includegraphics[width=85mm]{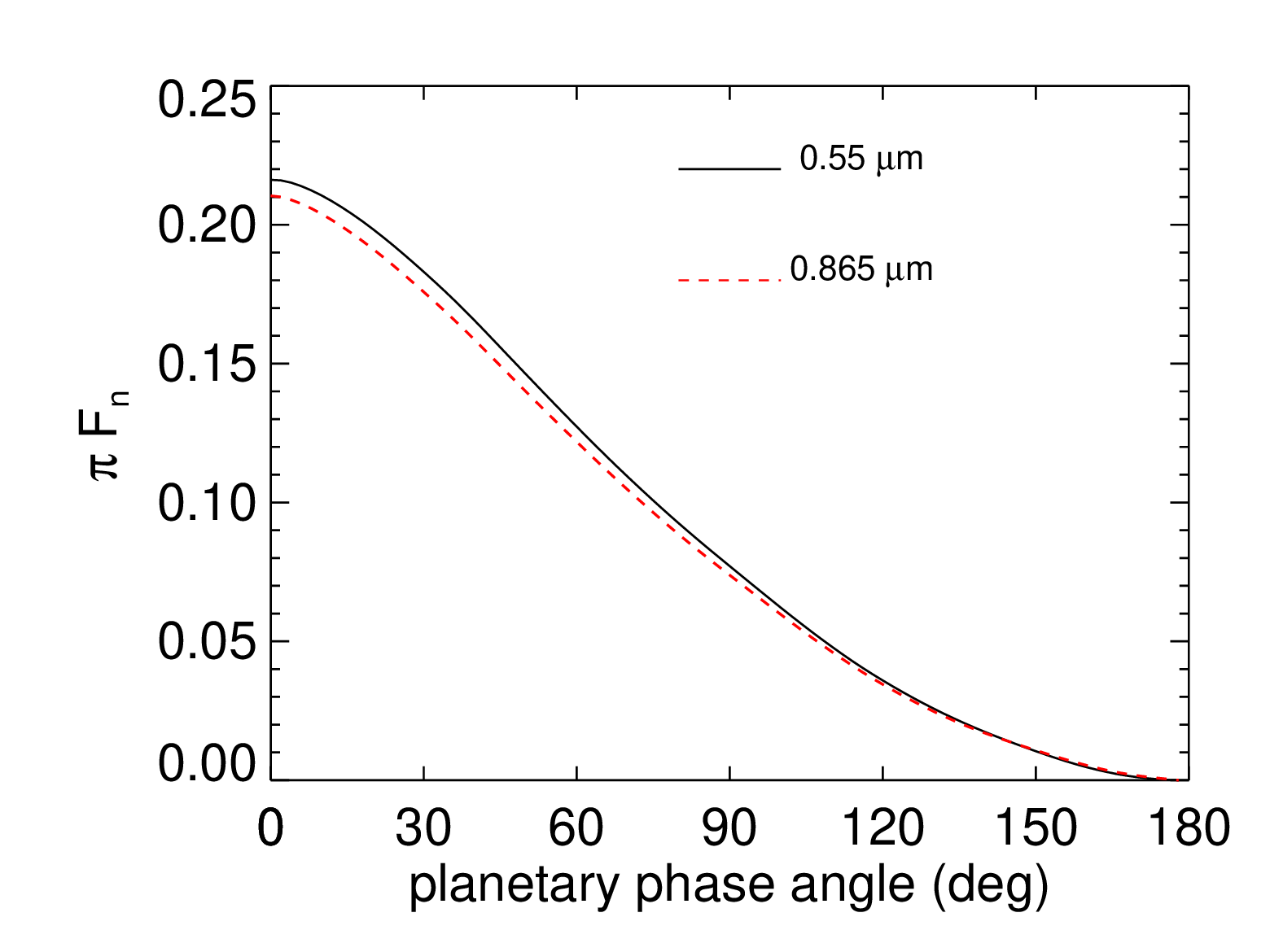}
\hspace{0.8cm}
\centering
\includegraphics[width=85mm]{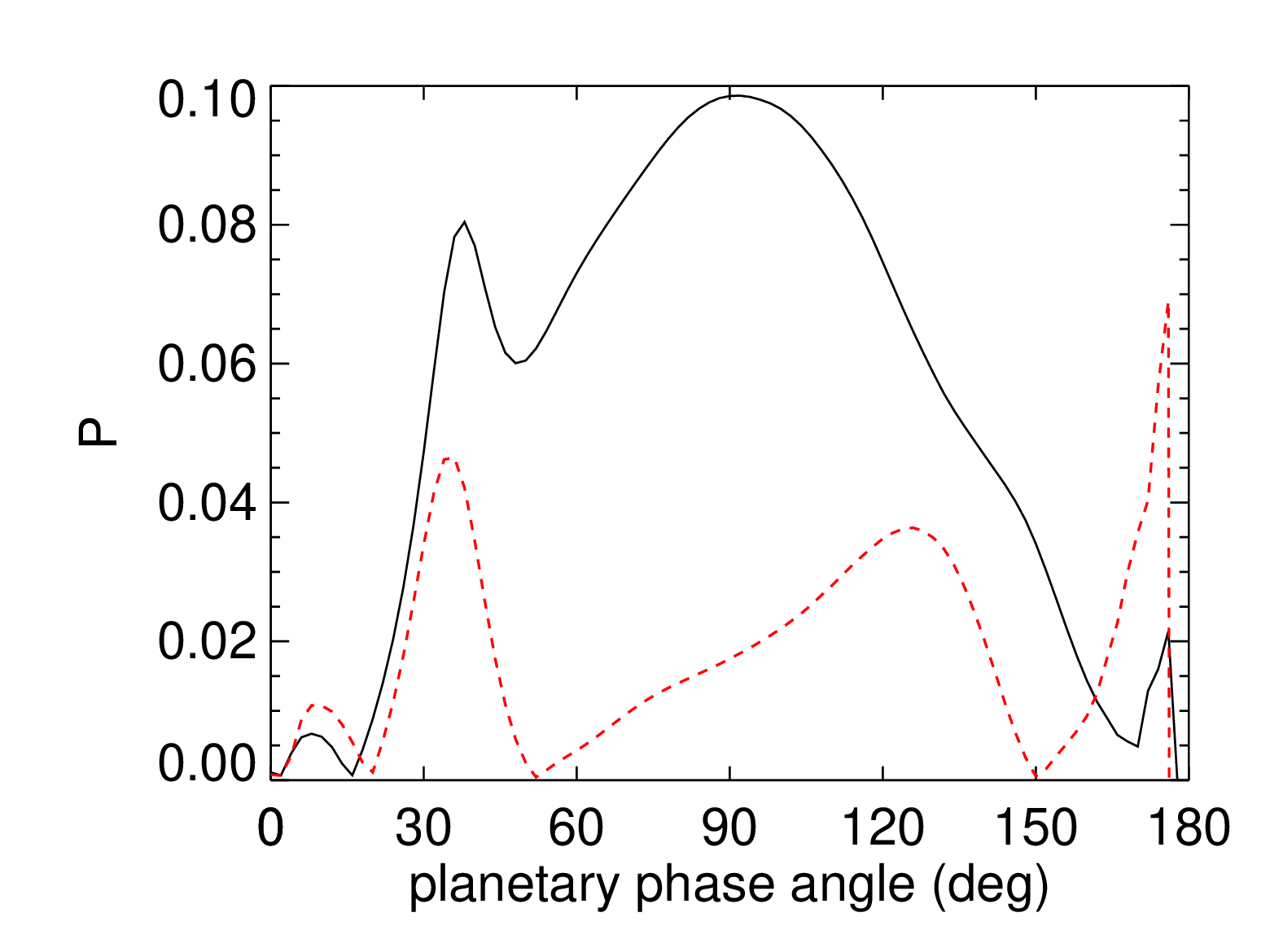}
\caption{$\pi F_\mathrm{n}$ and $P$ as functions of $\alpha$ for the
  model Earth with clouds as shown in Fig.~\ref{fig:modis_map}.}
\label{fig:modis_earth}
\end{figure}

In Fig.~\ref{fig:modis_earth}, we show the calculated $\pi
F_\mathrm{n}$ and $P$ of our model Earth at $\lambda=0.550~\mu$m and
0.865~$\mu$m.  The total reflected flux at $\alpha=0^\circ$ equals the
planet's geometric albedo, which is very similar at the two
wavelengths and equals $\sim$0.22 at $\lambda=0.55~\mu$m. This value
is slightly smaller than the one found in literature for Earth's
geometric albedo (0.33 \citep[][]{brown05}) and is due to our use of a
black surface on our planet.  It is clear that the total reflected
flux does not show a rainbow feature, while the degree of polarization
does. The maximum $P$ in the rainbow feature is 0.08 (8\% in
polarization) at $\lambda$=0.55~$\mu$m, and the absolute difference
with the nearby local minimum around $\alpha=50^\circ$, is 0.02 in $P$
(an absolute difference of 2\% in polarization). At
$\lambda=0.865~\mu$m, the rainbow feature is even more pronounced in
$P$, with an absolute difference with its surroundings of more than
4~$\%$. Interestingly, $P$ reaches zero around the rainbow feature
(which indicates a change of direction of the polarization), which
should facilitate measuring the strength of the rainbow feature.

\citet{bailey07} derived the disk-integrated polarization signature of
the partly-cloudy Earth from remote-sensing data and predicted degrees
of polarization of $\sim$12.7\% to $\sim$15.5\% in the rainbow peak at
wavelengths between about 0.5 to 0.8 microns. These values are higher
than our values of 4\% to 8\% as shown in
Fig.~\ref{fig:modis_earth}. The differences are, however, small when
considering the estimations by \citet{bailey07} on the influence of
multiple scattering and ice clouds on the polarization.

\section{Summary and conclusions}
\label{sect_summ}

The flux and degree of linear polarization of sunlight that is
scattered by cloud particles and that is reflected back to space
depend on the phase angle. Close to a phase angle of 40$^\circ$, the
flux and degree of polarization of unpolarized incident light that is
singly scattered by spherical water cloud droplets, shows the
enhancement that is known as the primary rainbow.  The multiple
scattering of light within clouds dilutes the primary rainbow feature
both in flux and in polarization.  In Earth--observation, with a
spatial resolution of a few kilometers, the detection of the polarized
rainbow feature is used to discriminate between liquid water clouds
and ice clouds \citep[][]{goloub00}. Light that is scattered by ice
cloud particles, such as hexagonal crystals, does not show the rainbow
feature. Knowledge of the cloud properties (thermodynamical phase,
optical thickness, microphysical properties etc) are crucial for
studies of global climate change on Earth \citep[][]{goloub00}.

In \citet{karalidi11}, we used horizontally homogeneous model planets
to investigate the strength of the rainbow feature in flux and
polarization for various model atmospheres of terrestrial
exoplanets. In that paper, we briefly discussed the influence of a
second atmospheric layer containing clouds on top of another one and
the implications that that could have on our ability to characterize
the planetary atmosphere. We concluded that even when the upper cloud
has a relatively small optical thickness, the polarization signal of
the planet is mainly determined by the properties of the upper cloud
particles. Here, in Sects.~\ref{sec:mixed_sizes} and~\ref{sec:h2oice}
we extended this research on horizontally inhomogeneous exoplanets and
for various cloud cases.

We noticed that in case the cloud layers contain clouds of similar
nature (for example liquid water clouds) and when the cloud particles
present a size stratification with altitude it is the top cloud layer
that will define the total planetary signal, as was also the case in
\citet{karalidi11}. On the other hand if the clouds are made out of
the same nature and size particles it is the (optically) thickest
clouds that will define to a largest extent the planetary signal.

In case the upper cloud layer contains ice clouds, we noticed that the
characteristics of the planetary signal can depend on either one of
the cloud layers, depending on their overcast and the optical
thickness of the ice cloud layer. Using our homogeneous planet code we
have seen that for $b_\mathrm{ice} \gtrsim 3$ the existence of any
lower cloud on the observed planetary pixel will be masked. When the
overcast of the two cloud layers is small the existence of the ice
cloud layer does not seem to be able to cover the existence of the
water cloud layer (i.e. the rainbow), even for high values of the
optical thickness (see Fig.~\ref{fig:icevar_b}).

Even in case the ice cloud layer has an optical thickness large enough
to mask the existence of the underlying liquid water cloud, the
rainbow feature of the liquid water cloud survives for the case the
ice cloud layer covers slightly more than half the water clouds (see
Fig.~\ref{fig:dbllr_h2o_ice_var_covs}). So unless our observed
exoplanet contains a very large number of ice clouds in its
atmosphere, the rainbow of the water clouds will still be visible in
the planetary $P$ signal.

An interesting test--case for the detection of the rainbow is
our own Earth. To test whether a distant observer would be able to 
detect a rainbow due to Earth's liquid water clouds partly covered
by water ice clouds, we modeled the Earth's cloud coverage using
MODIS/Aqua data from April 25, 2011. These data contained the 
location and optical thickness of ice and water clouds across the 
planet. We binned the cloud optical thicknesses in a limited number (5)
of optical thickness values (to avoid too many time--consuming computations),
as tabulated in Table~\ref{table:modis_model}, and modelled
the scattering properties of the liquid water cloud droplets using 
Mie-scattering and those of the water ice crystals using the 
models of \citet[][]{hess98}.

Our calculations for the disk--integrated flux and polarization
signals of this model Earth as functions of the planetary phase angle
and at $\lambda=0.550~\mu$m and 0.865~$\mu$m, show that the flux does
not have the primary rainbow feature. Flux observations as a function
of the phase angle would thus not provide an indication of the liquid
water clouds on Earth.  In the polarization signal, however, the
rainbow is clearly visible, especially at the longer wavelengths
($\lambda=0.865~\mu$m).  Polarimetry as a function of the planetary
phase angle would thus establish the existence of liquid water clouds
on our planet.

The results presented in this paper were for a black planetary
surface. A bright surface that reflects light with a low degree of
polarization would not significantly affect the reflected polarized
flux, but it would increase the total flux, and hence decrease the
degree of polarization. If a planet were thus covered by a bright
surface and few clouds, the rainbow feature would be less strong than
presented here (with increasing cloud coverage, the influence of the
surface would decrease), although P at other phase angles would also
be subdued. For example, comparing our results for a planet with a
mean Earth cloud coverage completely covered by a sandy surface with
an albedo of 0.243 (at 0.55 $\mu$m), polarization in the rainbow would
be about 15.39\%, compared to 41.7\% for a planet completely covered
by a black surface. Detailed calculations for planets with
realistically inhomogeneous surfaces and inhomogeneous cloud decks,
preferably including the variations of the cloud deck in time, would
help to study the effects of the surface reflection.  With an ocean
surface with waves, one could also expect the glint of starlight to
contribute to the reflected flux and polarization
\citep[][]{williams08}. How often this glint would be visible through
broken clouds and its effect on the rainbow of starlight that is
scattered by cloud particles, when it is indeed visible, will be
subject to further studies.

Summarizing, the primary rainbow of starlight that has been scattered
by liquid water clouds should be observable for modest coverages (10\%
- 20\%) of liquid clouds, and even when liquid water clouds are partly
covered by water ice clouds.  The total flux of the reflected
starlight as a function of the planetary phase angle does not show the
rainbow feature due to the presence of ice clouds.  The degree of
linear polarization of this light as a function of the phase angle
will usually show the rainbow feature, even when a large fraction (up
to $\sim$50\%) of the liquid water clouds are covered by ice clouds.
Polarimetry of starlight that is reflected by exoplanets, thus
provides a strong tool for the detection of liquid water clouds in the
planetary atmospheres.


\section*{Acknowledgments}

We would like to thank Dr. Michael Hess for providing us with the new
data set of the scattering properties of the ice crystals we use in
this paper. The authors acknowledge the MODIS mission scientists and
associated NASA personnel for the production of the data used in this
research effort.



\end{document}